\documentclass[journal,10pt,twocolumn]{IEEEtran}

\usepackage[T1]{fontenc}
\usepackage[utf8]{inputenc}
\usepackage{amsmath}
\usepackage{amssymb}
\usepackage{amsfonts}
\usepackage{mathtools}
\usepackage{amsthm}
\usepackage{microtype}
\usepackage{subfig}
\usepackage{booktabs}
\usepackage{tabularx}
\usepackage{multirow}
\usepackage{longtable}
\usepackage{float}
\usepackage{placeins}
\usepackage{array}
\usepackage{textcomp}
\usepackage{url}
\usepackage{verbatim}
\usepackage{graphicx}
\usepackage{xcolor}
\usepackage{color}
\usepackage{caption}
\usepackage{algorithm}
\usepackage{algorithmic}
\usepackage{tikz}
\usepackage{pgfplots}
\pgfplotsset{compat=1.18, every axis/.append style={font=\rmfamily, line width=0.8pt, tick style={line width=0.8pt}}}
\usetikzlibrary{positioning, shapes.geometric, arrows.meta, shapes.symbols, calc, fit, backgrounds, shadows.blur, chains, decorations.pathmorphing, shadows}
\tikzset{every node/.append style={font=\rmfamily}}
\usepackage{fontawesome5}
\usepackage{hyperref}
\hypersetup{hidelinks}

\definecolor{techBlue}{HTML}{2E5EAA}
\definecolor{techCyan}{HTML}{00A896}
\definecolor{techPurple}{HTML}{6D597A}
\definecolor{techOrange}{HTML}{F08A4B}
\definecolor{techGray}{HTML}{353535}
\definecolor{bgBlue}{HTML}{F0F4F8}
\definecolor{bgCyan}{HTML}{E6FFFA}
\definecolor{bgPurple}{HTML}{F3E5F5}
\definecolor{natureBlue}{HTML}{E1F5FE}
\definecolor{natureBlueDark}{HTML}{0277BD}
\definecolor{natureGreen}{HTML}{E8F5E9}
\definecolor{natureGreenDark}{HTML}{2E7D32}
\definecolor{naturePurple}{HTML}{F3E5F5}
\definecolor{naturePurpleDark}{HTML}{7B1FA2}
\definecolor{natureOrange}{HTML}{FFF3E0}
\definecolor{natureOrangeDark}{HTML}{EF6C00}
\definecolor{natureGray}{HTML}{F5F5F5}
\definecolor{natureGrayDark}{HTML}{424242}
\definecolor{ablationPurple}{HTML}{9EC4BE}
\definecolor{ablationYellow}{HTML}{F6C957}
\definecolor{ablationBlue}{HTML}{DCE9F4}
\definecolor{myColor1}{HTML}{FBE8D5}
\definecolor{myColor2}{HTML}{9EC4BE}
\definecolor{myColor3}{HTML}{ABD0F1}
\definecolor{myColor4}{HTML}{DCE9F4}
\definecolor{stepColor}{HTML}{DCE9F4}

\graphicspath{{./}}

\usepackage{stfloats}
\begin{document}

\title{GRAML: Graph-Grounded Reasoning and Multi-Task Learning for LLM-Based Software Vulnerability Detection}

\author{Ruohan Li,~Miaoqing Tian,~Haipeng Qu,~and~Ruobing Jiang%
\thanks{Ruohan Li and Miaoqing Tian are with Qingdao University of Technology, Qingdao, Shandong, China (e-mail: liruohan@stu.qut.edu.cn; tianmiaoqing@qut.edu.cn).}%
\thanks{Haipeng Qu and Ruobing Jiang are with Ocean University of China, Qingdao, Shandong, China (e-mail: quhaipeng@ouc.edu.cn; jrb@ouc.edu.cn).}%
\thanks{Corresponding author: Miaoqing Tian.}}

\maketitle

\begin{abstract}
Large Language Models (LLMs) have been widely applied to software vulnerability detection. However, their performance is often limited by insufficient use of control-flow and data-flow information. In this paper, we propose GRAML, a framework that combines graph evidence, vulnerability description generation, and multi-task training. GRAML first performs static analysis on C/C++ programs to extract critical source lines and typed line relations as structural evidence. It then uses this evidence to guide GPT-5 through the Tree-of-Thought-guided Vulnerability Reasoning (ToT-VR) process and generate vulnerability descriptions. These descriptions are further combined with Detection, Localization, and Assessment samples to build a unified four-task training dataset. We evaluate GRAML on an in-distribution (ID) test set and six out-of-distribution (OOD) datasets. The results show that GRAML achieves average F1 scores ranging from 66.67\% to 68.70\%, outperforming state-of-the-art baselines by up to 30.92\%. Ablation experiments further show that ToT-VR and graph-guided vulnerability descriptions improve detection performance compared with standard Chain-of-Thought (CoT) reasoning and raw Code Property Graph (CPG) serializations. These findings provide practical guidance for building more reliable and secure software engineering systems with large language models.
\end{abstract}


\section{Introduction}

With the rapid advancement of software engineering, LLMs have shown considerable promise in automating software vulnerability detection~\cite{ferrag2025securefalcon}. By learning from large volumes of source code and historical vulnerability samples, LLMs can capture general programming patterns and security-related semantics, enabling them to identify potential vulnerability patterns in complex codebases. These capabilities allow LLM-based approaches to reduce manual rule design and feature engineering, making them attractive for scalable vulnerability detection~\cite{fang2024large}. Recent studies further show that LLMs have been used in different vulnerability analysis settings, including software security surveys~\cite{sheng2025llms}, retrieval-based vulnerability detection~\cite{du2024vul}, and inter-procedural semantic analysis~\cite{wu2025enhancing}.

However, existing LLMs mainly understand source code from tokens and labels, but make limited use of control-flow and data-flow information~\cite{li2025sv,li2025iris}. As a result, they may miss how a vulnerability moves through the code from source to sink, making it difficult to identify its root cause~\cite{ding2024vulnerability}. This weakness can lead to both false positives, where benign code is misclassified as vulnerable, and false negatives, where real vulnerabilities are missed~\cite{li2025enhancing}. Existing studies have therefore explored additional program contexts and structural signals beyond binary labels, including vulnerability interpretation for LLM-based detection~\cite{du2024generalization}, CPG-guided context construction~\cite{lekssays2025llmxcpg}, and the fusion of language models and GNNs for joint semantic and structural modeling~\cite{liu2025vul}. Fine-grained property graph representations~\cite{shao2025graphfvd} and augmented Program Dependence Graph (PDG)-based representations~\cite{zou2025code} have also been used to preserve control-flow, data-flow, and semantic information in vulnerability detection. Nevertheless, how to extract and organize these structural relations into an effective context for downstream vulnerability reasoning remains an open problem.

\begin{figure*}[htbp]
    \centering
    \includegraphics[width=0.9\textwidth]{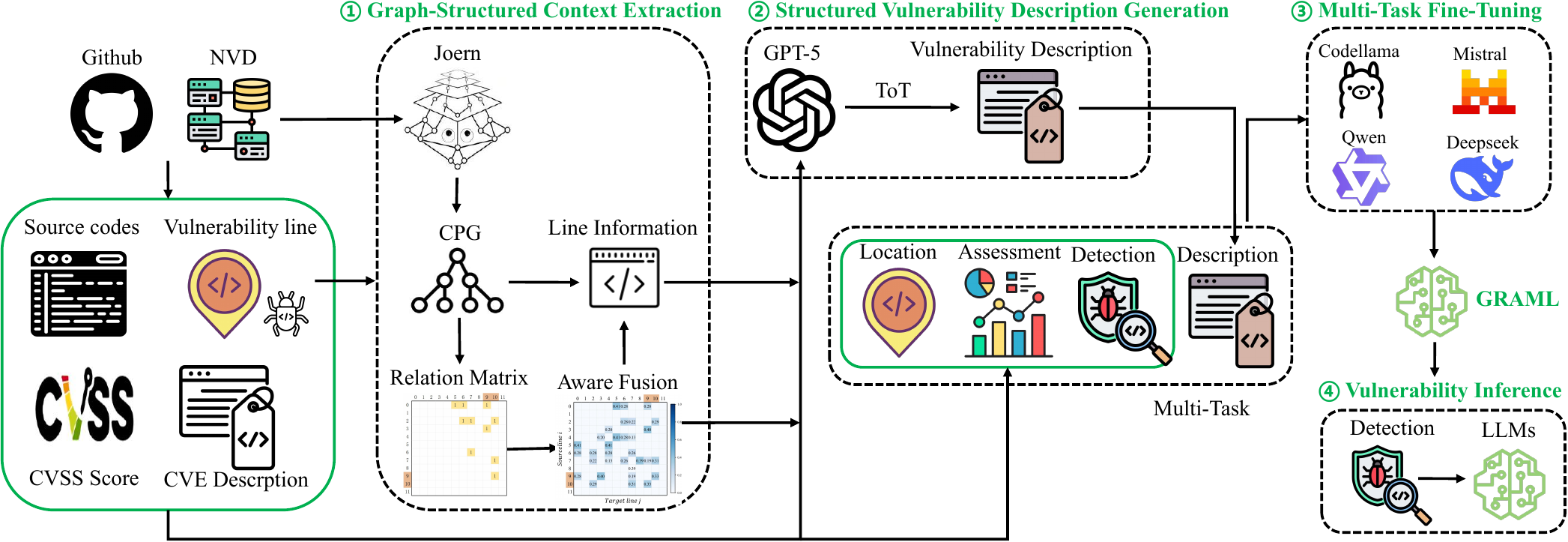}
    \caption{The Overview of GRAML. 
    }
    \label{fig:framework}
\end{figure*}

To address this challenge, we propose GRAML, a framework that integrates graph-based program information into LLM-based vulnerability detection. GRAML consists of graph evidence extraction, vulnerability description generation, and multi-task fine-tuning. During training data construction, we use Joern~\cite{joern2014home} to build Code Property Graphs (CPGs)~\cite{yamaguchi2014modeling}. We align graph nodes with physical source lines and retain data dependence, control dependence, and control flow relations. Starting from known vulnerable lines obtained from patch information or CVE records, GRAML applies reliability-aware propagation to select critical lines and typed line relations. The resulting graph evidence is then used to support vulnerability description generation.

Based on the extracted evidence, we design ToT-VR to generate vulnerability descriptions with GPT-5~\cite{openai2025gpt5}. The reasoning process uses the source code and available CVE information together with the selected critical lines and graph relations. It first constructs multiple candidate analyses. These candidates are then pruned, refined, and checked against the available evidence. The final output describes the vulnerability cause, the affected code regions, and the potential security consequences.

We then construct a training dataset using BigVul~\cite{fan2020ac}, DiverseVul~\cite{chen2023diversevul}, and newly collected CVE samples reported between 2019 and 2026. The dataset contains four tasks, namely Detection, Description, Localization, and Assessment. Detection determines whether the input code is vulnerable. Description explains the vulnerability cause, Localization identifies the affected code lines, and Assessment estimates the vulnerability severity. The descriptions generated by ToT-VR are used for the Description task. All four tasks are combined to train the models jointly.

Using this multi-task dataset, we fine-tune CodeLlama, Mistral, DeepSeek, and Qwen in the GRAML framework. After fine-tuning, each model requires only source code during inference and performs the Detection task without graph evidence, vulnerable-line annotations, or CVE descriptions. We evaluate the four models on the in-distribution (ID) test set and six out-of-distribution (OOD) datasets. The experimental results show that their average F1 scores range from 66.67\% to 68.70\%, outperforming state-of-the-art baselines by up to 30.92\%. Although these models use 6.7B--7B parameter backbones, they achieve higher average F1 scores than the larger general-purpose LLMs included in our evaluation. To provide a fair comparison with VulLLM~\cite{du2024generalization}, we replace GPT-5 with GPT-4 in GRAML while keeping the remaining settings unchanged. Under this setting, all four GRAML models achieve higher average F1 scores than the two VulLLM models.
We further compare ToT-VR with standard Chain-of-Thought (CoT) and examine how graph evidence and its representation affect detection performance. We also evaluate model robustness under code transformations that preserve program behavior.

The main contributions of this paper are summarized as follows:

\begin{itemize}

    \item \textbf{Reliability-aware graph context extraction.}
    We construct a sparse line-level CPG representation with typed data-dependence, control-dependence, and control-flow relations. The method combines vulnerable-line reliability, degree-normalized propagation, relation diversity, and redundancy control to select compact evidence for vulnerability reasoning.

    \item \textbf{Graph-guided vulnerability reasoning with ToT-VR.} We design a five-stage ToT-VR process based on GPT-5. It constructs multiple reasoning branches, removes unsupported analyses, refines the remaining candidates, verifies them against the available evidence, and generates the final vulnerability description. This process converts the extracted graph evidence into natural-language descriptions that are used for model training.

    \item\textbf{Expanded multi-task vulnerability dataset.}
    We construct a vulnerability training dataset by combining BigVul~\cite{fan2020ac}, DiverseVul~\cite{chen2023diversevul}, and newly collected CVE samples reported between 2019 and 2026.  We organize the data into four tasks, including Detection, Description, Localization, and Assessment. These tasks provide supervision from vulnerability labels, causes, affected locations, and severity information to improve vulnerability detection.

\end{itemize}

The remainder of this paper is structured as follows. We first present the proposed GRAML framework and its key components in Section \ref{sec:methodology}. We then describe the experimental setup and analyze the results in Section \ref{sec:evaluation}. Related work is reviewed in Section \ref{sec:related_work}, followed by a discussion of the implications of our findings and potential limitations in Section \ref{sec:discussion}. Finally, Section \ref{sec:conclusion} concludes the paper and highlights future research directions.

\section{Methodology}
\label{sec:methodology}

\begin{figure*}[htbp]
    \centering
    \includegraphics[width=\textwidth]{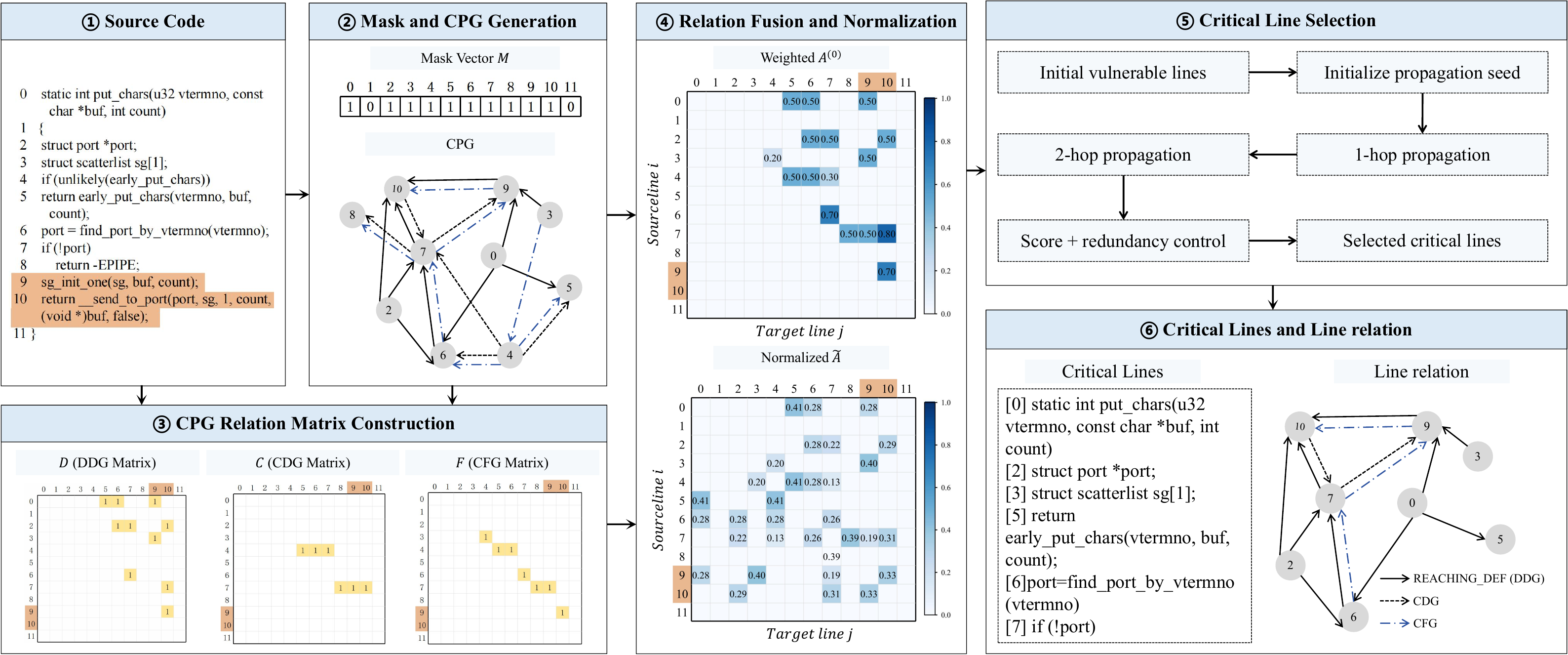}
    \caption{Graph-structured context extraction: CPG generation and line alignment, reliability-aware multi-relation matrix construction, and vulnerability-line-aware critical evidence selection.}
    \label{fig:combined_extraction_process}
\end{figure*}

\subsection{Overview}
We propose GRAML, a framework that integrates graph-structured program analysis with multi-task training to enhance the vulnerability detection performance of LLMs. The overview of GRAML is illustrated in Figure \ref{fig:framework}. 
The framework consists of four sequential modules, including (1) \textbf{Graph-Structured Context Extraction}. This module uses Joern to construct CPGs from the source code. It then identifies critical lines based on the vulnerable lines and extracts their structural dependencies to build graph-structured contexts (Section \ref{sec:mask_construction});(2) \textbf{Structured Vulnerability Description Generation}, which combines the graph-structured contexts, source code, and CVE descriptions in a ToT-VR process to generate structured vulnerability description (Section \ref{sec:description_generation}); (3) \textbf{Multi-Task Fine-Tuning}, which utilizes the extracted information from previous steps to train LLMs with four tasks, including vulnerability detection, description, localization, and assessment (Section \ref{sec:multitask_tuning}); and (4) \textbf{Vulnerability Inference}, which applies the fine-tuned LLMs to detect vulnerabilities and evaluate their detection performance (Section \ref{sec:inference_strategy}).

\subsection{Graph-Structured Context Extraction}
\label{sec:mask_construction}

The graph-structured context serves only as an intermediate representation for generating vulnerability descriptions. It is not directly included in the input to the fine-tuned LLM. As illustrated in Figure~\ref{fig:combined_extraction_process}, the extraction process comprises six components. In Step~1, the source-code function is indexed by its physical line numbers, and the vulnerable lines are marked. In Step~2, a binary mask and a line-aligned CPG are generated. In Step~3, the Data Dependence Graph (DDG), Control Dependence Graph (CDG), and Control Flow Graph (CFG) relations extracted from the CPG are represented by three relation-specific matrices. In Step~4, these matrices are fused into a weighted matrix and normalized into a structural relation matrix. In Step~5, the normalized matrix is used to select critical contextual lines related to the vulnerable lines. Finally, Step~6 retains the selected lines and their directed relations as the graph-structured context. The following paragraphs describe each step in detail.

\textbf{Source Code.}
The input is a vulnerable function with its original physical line numbers. Before graph-structured context extraction, we compare the vulnerable function with its patched version. The changed lines associated with the vulnerability are identified as vulnerable lines. Their physical line numbers are retained as vulnerable-line annotations and are used consistently in the subsequent mask, CPG, relation matrices, and critical-line selection.

As shown in Figure~\ref{fig:combined_extraction_process}, we use the \texttt{put\_chars()} function associated with CVE-2017-8067 as an example. The function contains 12 physical lines, numbered from 0 to 11. By comparing the vulnerable function with its patched version, lines 9 and 10 are identified as the vulnerable lines. These lines contain calls to \texttt{sg\_init\_one()} and \texttt{\_\_send\_to\_port()}, respectively. They are highlighted throughout the following steps so that their roles in the extraction process can be clearly tracked.

\textbf{Mask and CPG Generation.}
Given a function with $N$ physical lines, we first construct a binary mask $M \in \{0,1\}^{N}$. We set $M_i=1$ for a non-empty and non-comment line and $M_i=0$ otherwise. We then use Joern to construct a CPG from the source code and map its nodes to the corresponding physical source lines. The CPG provides three types of directed line-level relations, namely the Data Dependence Graph (\textsc{DDG}), Control Dependence Graph (\textsc{CDG}), and Control Flow Graph (\textsc{CFG}) relations. The DDG relations, represented by REACHING\_DEF edges, describe how definitions of variables reach their subsequent uses. The CDG relations, represented by CDG edges, describe how the execution of statements depends on control conditions. The CFG relations, represented by CFG edges, describe the possible execution order between statements. These relations provide the structural information used to construct the subsequent relation matrices.

In the example shown in Figure~\ref{fig:combined_extraction_process}, lines 1 and 11 contain only braces and are therefore assigned 0, while the remaining lines are assigned 1. The highlighted indices 9 and 10 mark the vulnerable lines and do not change their mask values. In the line-aligned CPG, the declaration of \texttt{sg} at line 3 is connected to its use at line 9 through a DDG relation. The value initialized at line 9 is subsequently used at line 10, forming another DDG relation. The condition at line 7 controls the execution of lines 8, 9, and 10, resulting in CDG relations. The CFG relations record execution transitions such as those from line 6 to line 7 and from line 9 to line 10.

\textbf{CPG Relation Matrix Construction.}
Using the line-aligned CPG obtained in Step~2, we construct three binary matrices for the DDG, CDG, and CFG relations. The rows and columns of each matrix correspond to the physical source-line numbers. Let $D_{ij}$, $C_{ij}$, and $F_{ij}$ indicate whether a directed DDG, CDG, or CFG relation exists from source line $i$ to source line $j$, respectively. An entry is set to 1 when the corresponding relation exists and to 0 otherwise.

As shown in Step~3 of Figure~\ref{fig:combined_extraction_process}, the line-aligned CPG is converted into the DDG matrix $D$, the CDG matrix $C$, and the CFG matrix $F$. Rows and columns 9 and 10 are highlighted because they correspond to the vulnerable lines. For example, $D_{3,9}=1$ records the DDG relation from line 3 to line 9, $C_{7,10}=1$ records the CDG relation from line 7 to line 10, and $F_{9,10}=1$ records the CFG relation from line 9 to line 10. These three binary matrices are passed to Step~4 for relation fusion and normalization.

\textbf{Relation Fusion and Normalization.}
The three binary matrices constructed in Step~3 record the DDG, CDG, and CFG relations separately. In Step~4, we combine these matrices into a weighted directed relation matrix. The mask $M$ generated in Step~2 is also used to exclude relations involving empty or comment-only lines. The weighted matrix is defined as
\begin{equation}
A^{(0)}_{ij}=M_iM_j\left(0.50D_{ij}+0.30C_{ij}+0.20F_{ij}\right).
\label{eq:relation_matrix}
\end{equation}
Here, $D_{ij}$, $C_{ij}$, and $F_{ij}$ denote the entries of these three matrices. The DDG, CDG, and CFG weights are set to $0.50$, $0.30$, and $0.20$, respectively, based on validation experiments. This setting assigns the largest weight to data dependence while retaining control dependence and execution order. The same weights are used for all samples.

As shown in the upper matrix of Step~4 in Figure~\ref{fig:combined_extraction_process}, $A^{(0)}$ combines the weights of all relation types between the same pair of lines. If multiple relation types exist, their weights are added. For example, the relation from line 4 to line 5 contains both CDG and CFG information and therefore receives a weight of $0.50$. The relation from line 6 to line 7 contains both DDG and CFG information and receives a weight of $0.70$. Similarly, the relation from line 7 to vulnerable line 10 contains both DDG and CDG information and receives a weight of $0.80$, while the relation from vulnerable line 9 to vulnerable line 10 contains both DDG and CFG information and receives a weight of $0.70$. These values are recorded as the corresponding entries in the weighted directed matrix $A^{(0)}$.

The matrix $A^{(0)}$ preserves the direction of each relation. However, critical-line selection must consider code lines connected to vulnerable lines through both incoming and outgoing relations. We therefore combine $A^{(0)}$ with its transpose:
\begin{equation}
B=A^{(0)}+(A^{(0)})^{\top}.
\label{eq:support_matrix}
\end{equation}
The resulting matrix $B$ records the connection between two lines regardless of its original direction. However, code lines may have different numbers of structural connections. Without normalization, a line with many connections may receive a large value simply because it has more neighbors. To reduce this effect, we normalize $B$ as follows:
\begin{equation}
\widetilde{A}=\Delta^{-\frac{1}{2}}B\Delta^{-\frac{1}{2}}, \qquad \Delta_{ii}=\sum_j B_{ij},
\label{eq:normalized_relation_matrix}
\end{equation}
where $\Delta_{ii}$ is the total relation weight connected to line $i$. This normalization reduces the influence of highly connected lines.

As shown in the lower matrix of Step~4 in Figure~\ref{fig:combined_extraction_process}, the directed DDG relation from line 3 to vulnerable line 9 has a weight of $0.50$ in $A^{(0)}$. After combining $A^{(0)}$ with its transpose, the connection is recorded at both $(3,9)$ and $(9,3)$. After normalization, both entries have a value of approximately $0.40$ in $\widetilde{A}$.

Step~4 therefore produces two matrices with different functions. The weighted matrix $A^{(0)}$ preserves the direction, type, and weight of each CPG relation and is retained for the final relation records. The normalized matrix $\widetilde{A}$ describes the structural connections around each line and is passed to Step~5 for critical-line selection.

\textbf{Critical Line Selection.}
Step~5 starts from the vulnerable-line set $S_{vul}$ identified in Step~1. We convert these vulnerable lines into an initial vector $z^{(0)}$:
\begin{equation}
z^{(0)}_i=
\begin{cases}
\alpha_i / \sum_{v\in S_{vul}}\alpha_v, & i\in S_{vul},\\
0, & \text{otherwise}.
\end{cases}
\label{eq:vulnerable_line_initialization}
\end{equation}
Here, $\alpha_i$ denotes the reliability assigned to vulnerable line $i$, and all vulnerable lines in $S_{vul}$ are assigned $\alpha_i=1$. The denominator $\sum_{v\in S_{vul}}\alpha_v$ is the total reliability accumulated over all vulnerable lines of the current function, so the normalization distributes a total weight of $1$ among them; the remaining lines are assigned 0. In the example function, $S_{vul}=\{9,10\}$, which are its only vulnerable lines, and $\alpha_9=\alpha_{10}=1$; the two vulnerable lines therefore share the unit weight equally, yielding $z^{(0)}_9=z^{(0)}_{10}=0.5$ and $z^{(0)}_i=0$ for all other lines.

We then use the normalized matrix $\widetilde{A}$ obtained in Step~4 to calculate the structural relevance of other code lines. The one-hop and two-hop relevance vectors are defined as
\begin{equation}
z^{(1)}=\widetilde{A}z^{(0)},
\label{eq:first_hop_propagation}
\end{equation}
and
\begin{equation}
z^{(2)}=\widetilde{A}z^{(1)}.
\label{eq:second_hop_propagation}
\end{equation}
The vector $z^{(1)}$ represents the relevance of lines directly connected to the vulnerable lines, whereas $z^{(2)}$ represents the relevance of lines connected through one intermediate code line.

To select the most relevant candidates, we calculate a score for each line:
\begin{equation}
\operatorname{Score}(i)=0.45\widehat{z}^{(1)}_i
+0.25\widehat{z}^{(2)}_i
+0.20d_i
-0.10\rho_i,
\label{eq:critical_score}
\end{equation}
where $\widehat{z}^{(1)}_i$ and $\widehat{z}^{(2)}_i$ are the normalized one-hop and two-hop relevance values. Let $\mathcal{T}_i$ denote the set of relation types among DDG, CDG, and CFG that connect candidate line $i$ to the currently selected lines. We define
\begin{equation}
d_i=\frac{|\mathcal{T}_i|}{3}.
\label{eq:relation_diversity}
\end{equation}
Thus, $d_i$ is larger when the candidate is connected through more relation types.

Let $\mathcal{N}(i)$ denote the set of code lines connected to line $i$ through DDG, CDG, or CFG relations, and let $S_{\mathrm{sel}}$ denote the set of previously selected lines. We define the redundancy of candidate line $i$ as
\begin{equation}
\rho_i=
\max_{j\in S_{\mathrm{sel}}}
\frac{|\mathcal{N}(i)\cap\mathcal{N}(j)|}
     {|\mathcal{N}(i)\cup\mathcal{N}(j)|}.
\label{eq:structural_redundancy}
\end{equation}
When $S_{\mathrm{sel}}$ is empty, we set $\rho_i=0$. A large $\rho_i$ indicates that the candidate has a structural neighborhood similar to that of an already selected line. Therefore, the first three terms favor lines that are close to the vulnerable lines and contain diverse structural relations, while the final term reduces the selection of redundant lines.

As shown in Step~5 of Figure~\ref{fig:combined_extraction_process}, the selection starts from vulnerable lines 9 and 10. The one-hop and two-hop calculations identify structurally related candidates, and the scoring process selects contextual lines 0, 2, 3, 5, 6, and 7 after removing structurally repetitive candidates. Vulnerable lines 9 and 10 are also retained to preserve their relations with the selected context. If the CPG contains no usable relation, the evidence contains only the vulnerable lines. Finally, we retain the directed DDG, CDG, and CFG relations in $A^{(0)}$ whose source and destination lines both belong to $S_{crit}$.

\textbf{Critical Lines and Line Relations.}
Step~6 outputs the selected critical lines and the relations between them. For each critical line, we retain its physical line number and source-code text. For each relation, we retain its source line, destination line, direction, and relation type. These records form the final graph-structured context.

As shown in Step~6 of Figure~\ref{fig:combined_extraction_process}, the selected contextual lines are 0, 2, 3, 5, 6, and 7. Vulnerable lines 9 and 10 are also retained to preserve their connections with the contextual lines. The arrows in the figure show the relations among these lines. Solid black arrows denote DDG relations. Dashed arrows denote CDG relations. Dash-dotted arrows denote CFG relations. These critical lines and directed relations are used in the following ToT-VR stage to generate vulnerability descriptions.

\begin{figure}[htbp]
    \centering
    \includegraphics[width=1.0\columnwidth]{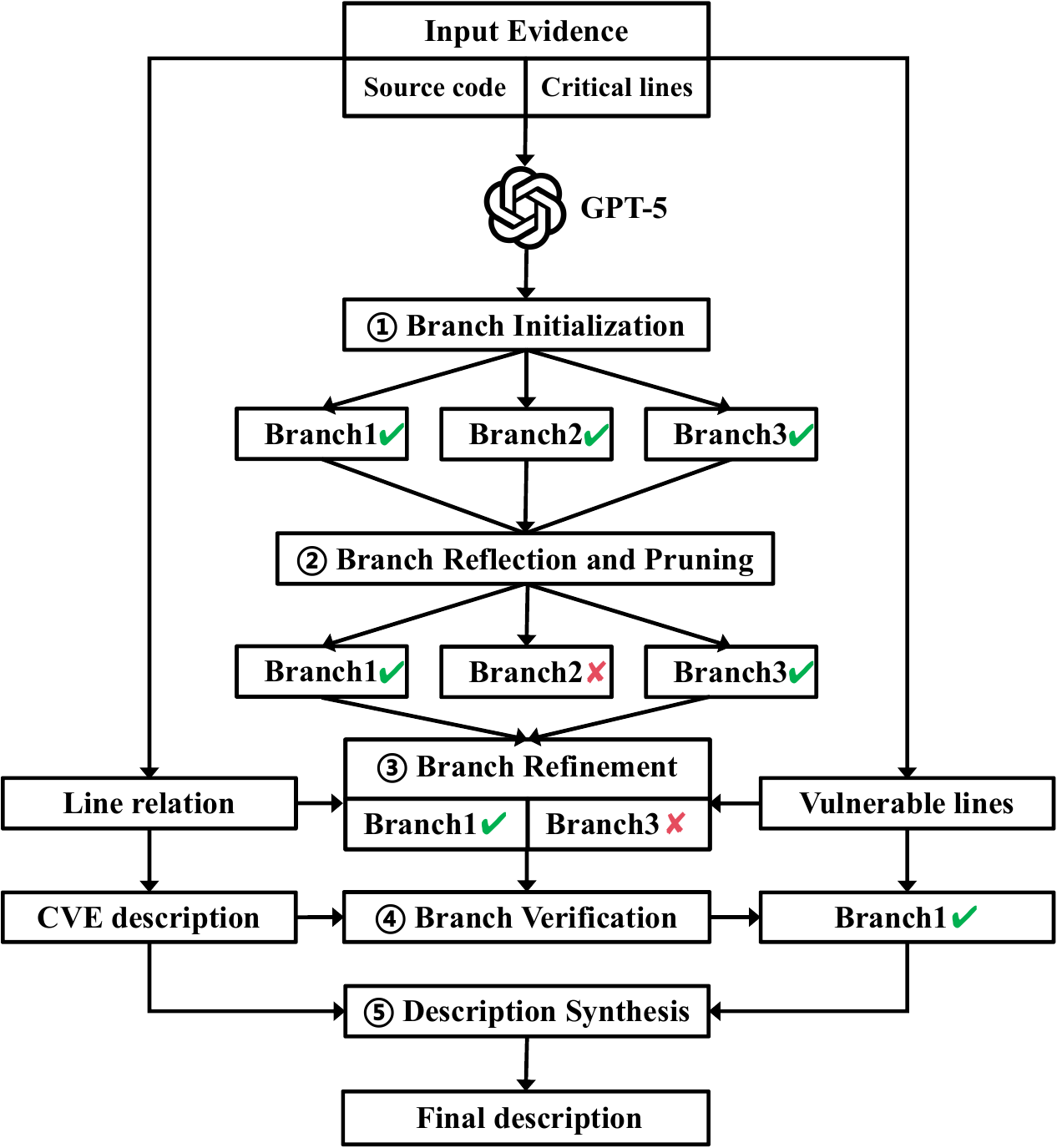}
    \caption{Overview of the proposed ToT-VR process.}
    \label{fig:tot_process}
\end{figure}

\subsection{Structured Vulnerability Description Generation}
\label{sec:description_generation}

Tree of Thoughts (ToT) \cite{yao2023tree} extends conventional CoT reasoning by exploring multiple candidate reasoning paths and selecting the most promising ones through intermediate evaluation. We use GPT-5 to perform structured vulnerability reasoning and propose a \textbf{Tree-of-Thought-guided Vulnerability Reasoning (ToT-VR)} mechanism for generating vulnerability descriptions. In ToT-VR, GPT-5 receives the source code and up to 20 critical lines as initial structural evidence. The selected vulnerable lines and typed line relations are then introduced during branch refinement, and the official CVE description is provided during branch verification. This evidence guides multi-branch reasoning, but it is not directly serialized into the downstream fine-tuning input. Each branch corresponds to a candidate vulnerability interpretation that focuses on a trigger point, dependency path, or causal explanation. The process proceeds stage by stage. Figure \ref{fig:tot_process} presents the overall workflow, and Table \ref{tab:tot_branch_example} tracks a running example through the stages. The remainder of this subsection describes each stage in turn.

\begin{table}[t]
\centering
\caption{An example illustrating the main stages of ToT-VR, where B1, B2, and B3 denote three candidate reasoning branches generated during branch initialization.}
\label{tab:tot_branch_example}
\resizebox{\columnwidth}{!}{%
\begin{tabular}{@{}l p{0.76\columnwidth}@{}}
\toprule
\textbf{Stage} & \textbf{Summary} \\
\midrule

\textbf{Code Snippet}
& \texttt{char tmp[64]; strcpy(tmp, buf); printf("\%s", tmp);} \\
\midrule

\textbf{Branch Initialization}
& \textbf{B1:} stack buffer overflow from unchecked \texttt{strcpy}. \newline
  \textbf{B2:} format string vulnerability related to \texttt{printf}. \newline
  \textbf{B3:} incorrect input-length handling. \\
\midrule

\textbf{Branch Reflection and Pruning}
& Keep \textbf{B1} and \textbf{B3}; prune \textbf{B2} because \texttt{printf("\%s", tmp)} uses a constant format string. \\
\midrule

\textbf{Branch Refinement}
& Incorporate the selected vulnerable lines and typed line relations. Retain \textbf{B1}; prune \textbf{B3} because the structural evidence does not support an input-length-handling issue. \\
\midrule

\textbf{Branch Verification}
& \textbf{B1} is verified as it best matches the CVE description and known vulnerability semantics. \\
\midrule

\textbf{Description Synthesis}
& Generate a unified description covering the vulnerability cause, the affected code regions, and the potential security consequences. \\
\midrule

\textbf{Final Description}
& Stack-based buffer overflow caused by unbounded \texttt{strcpy} into the fixed-size buffer \texttt{tmp}. \\
\bottomrule
\end{tabular}%
}
\end{table}

\begin{itemize}
    \item \textbf{Step 1: Branch Initialization.} Given the source code and up to 20 critical lines extracted in the previous stage, GPT-5 first constructs multiple candidate reasoning branches. As shown in Table \ref{tab:tot_branch_example}, different branches may emphasize different trigger points or causal mechanisms. For example, the model may interpret the code as a stack buffer overflow caused by unchecked copying, a format string vulnerability related to \texttt{printf}, or an issue caused by incorrect handling of input length.

    \item \textbf{Step 2: Branch Reflection and Pruning.} GPT-5 then reviews the generated branches for internal consistency and alignment with the code semantics. Based on this review, it revises weak branches and removes unsupported ones, retaining only the branches that provide reasonable descriptions of the potential vulnerability. This step filters out noisy or contradictory reasoning branches before additional structural evidence is introduced. As shown in Table \ref{tab:tot_branch_example}, this stage retains Branch1 and Branch3 while pruning Branch2.

    \item \textbf{Step 3: Branch Refinement.} To further ground the reasoning process, we provide the selected vulnerable lines and the typed line relations. GPT-5 uses this evidence to refine the retained branches and align them with more precise trigger points and dependency paths. As shown in Table \ref{tab:tot_branch_example}, this stage retains Branch1 and prunes Branch3. In this way, graph-derived program structure constrains the reasoning process without requiring the fine-tuned model to interpret a raw graph serialization.

    \item \textbf{Step 4: Branch Verification.} We further provide the official CVE description as external semantic evidence. At this stage, GPT-5 compares the retained branches against the CVE description and evaluates whether they are consistent with the known vulnerability semantics. If the retained branches are not supported by the CVE description, the model can return to the candidate branches and reconsider alternative interpretations. This step serves as the final semantic verification stage to ensure that the retained reasoning is not only structurally grounded but also aligned with real-world vulnerability descriptions. In the example shown in Table \ref{tab:tot_branch_example}, Branch1 is finally retained because it is consistent with both the structural evidence and the official CVE description.

    \item \textbf{Step 5: Description Synthesis.} Finally, GPT-5 synthesizes the retained branch into a complete natural-language vulnerability description. The description covers the vulnerability cause, the affected code regions, and the potential security consequences. This explanation, rather than the raw CPG evidence, serves as the supervision target for the Description task. The complete prompt templates used in all five stages of ToT-VR are provided in Appendix~\ref{sec:appendix_tot_prompts}.

\end{itemize}

To implement the ToT-VR process described above, we use GPT-5 through the OpenAI API. A Python script controls the five reasoning stages for each vulnerable function. It first prepares the source code, the selected vulnerable lines, up to 20 critical lines, and the typed line relations between these lines. Each critical line contains only its physical line number and source-code text. The source code and critical lines are used to initialize candidate branches. The selected vulnerable lines and typed line relations are then used to refine the trigger points, dependency paths, and causal descriptions of the retained branches. The official CVE description is provided only during branch verification. After verification, the retained branch is converted into a natural-language vulnerability description. This description serves as the target of the Description task during multi-task fine-tuning. The intermediate branches and structural records are stored for traceability, but they are not included in the final fine-tuning input. The script also handles temporary API errors and checks the output format to maintain consistent results.

\begin{table}[t]
\centering
\caption{Target output formats for the four fine-tuning tasks.}
\label{tab:output_formats}
\scriptsize
\renewcommand{\arraystretch}{1.18}
\begin{tabularx}{\columnwidth}{
>{\raggedright\arraybackslash}p{0.18\columnwidth}
>{\raggedright\arraybackslash}p{0.39\columnwidth}
>{\raggedright\arraybackslash}X
}
\toprule
\textbf{Task} &
\textbf{Target for Vulnerable Samples} &
\textbf{Target for Non-Vulnerable Samples} \\
\midrule

\textbf{Detection}
& ``Yes'' (mapped to 1)
& ``No'' (mapped to 0) \\

\midrule

\textbf{Description}
& A paragraph describing the vulnerability cause, the affected code regions, and the potential security consequences.
& ``There is no vulnerability.'' \\

\midrule

\textbf{Localization}
& Comma-separated physical line indices, e.g., ``1, 4, 7''.
& ``N/A'' \\

\midrule

\textbf{Assessment}
& A CVSS v3.0 severity class, i.e., ``0'', ``1'', or ``2''.
& ``N/A'' \\

\bottomrule
\end{tabularx}
\end{table}

\begin{table}[t]
\centering
\caption{CVSS v3.0 severity mapping used in the Assessment task.}
\label{tab:cvss_mapping}
\scriptsize
\renewcommand{\arraystretch}{1.18}
\begin{tabularx}{\columnwidth}{
>{\raggedright\arraybackslash}p{0.32\columnwidth}
>{\raggedright\arraybackslash}p{0.15\columnwidth}
>{\raggedright\arraybackslash}X
}
\toprule
\textbf{CVSS v3.0 Score} & \textbf{Output} & \textbf{Severity Class} \\
\midrule
$7.0 \leq s \leq 10.0$ & ``2'' & High/Critical \\
\midrule
$4.0 \leq s < 7.0$ & ``1'' & Medium \\
\midrule
$0.0 < s < 4.0$ & ``0'' & Low \\
\midrule
Non-vulnerable sample & ``N/A'' & --- \\
\bottomrule
\end{tabularx}
\end{table}

\subsection{Multi-Task Fine-Tuning}
\label{sec:multitask_tuning}

We construct the training dataset from BigVul~\cite{fan2020ac}, DiverseVul~\cite{chen2023diversevul}, and additional vulnerable functions collected from CVE reports published between 2019 and 2026. Based on these samples, we formulate four tasks, including Detection, Description, Localization, and Assessment. The \textbf{Detection} task predicts whether a function is vulnerable. Given a source function, the model outputs ``Yes'' or ``No'' according to its vulnerability label. The \textbf{Localization} task identifies the vulnerable lines in a function. Vulnerable samples use the vulnerable-line annotations described in Section~\ref{sec:mask_construction}, while nonvulnerable samples use \texttt{N/A}. The \textbf{Assessment} task predicts the vulnerability severity. Vulnerable samples use the severity level derived from the CVSS v3.0 score, while nonvulnerable samples use \texttt{N/A}. The \textbf{Description} task generates a vulnerability explanation. Vulnerable samples use the descriptions generated by ToT-VR in Section~\ref{sec:description_generation}, while nonvulnerable samples use a response indicating that no vulnerability exists. Detection is the primary task, while Description, Localization, and Assessment provide additional information about the vulnerability cause, affected lines, and severity. Table~\ref{tab:output_formats} summarizes the target formats of each task, and Table~\ref{tab:cvss_mapping} presents the severity mapping used for Assessment. More details about dataset construction and task distribution are provided in Section~\ref{sec:datasets}.

We organize all tasks using a unified instruction format. Each training sample contains three components, including a task instruction, a source-code input, and a target response. The task instruction defines the objective of the task, the source-code input provides the function to be analyzed, and the target response specifies the expected output. The complete prompts for Detection, Description, Localization, and Assessment are provided in Appendix~\ref{app:task_instruction_prompts}.

We fine-tune four LLMs, including Qwen2.5-Coder-7B-Instruct~\cite{hui2024qwen2}, CodeLlama-7B-Instruct-hf~\cite{roziere2023code}, Mistral-7B-Instruct-v0.2~\cite{jiang2023mistral}, and DeepSeek-Coder-6.7B-Instruct~\cite{guo2024deepseek}. Since these models adopt different chat templates, we convert each training sample into the format required by the corresponding model. The task instruction, source code, and target response remain unchanged, while the message structure and special tokens are adjusted for each model. The complete model-specific chat templates are provided in Appendix~\ref{app:target_llm_prompts}. During fine-tuning, we optimize the models with a causal language modeling objective and apply completion-only masking. Specifically, the task instruction and source-code tokens are excluded from loss computation, and only the target-response tokens are used for optimization.

We implement all experiments using PyTorch~\cite{paszke2019pytorch}, Hugging Face Transformers~\cite{wolf2020transformers}, and Unsloth~\cite{unsloth}. We conduct all experiments on six NVIDIA RTX 5090 GPUs, each with 32\,GB of memory. We use QLoRA with 4-bit quantization and set the LoRA rank to $r=32$ and the scaling factor to $\alpha=16$. LoRA adapters are applied to all linear projection modules in the Transformer layers, including attention projections (\texttt{q\_proj}, \texttt{k\_proj}, \texttt{v\_proj}, and \texttt{o\_proj}) and feed-forward projections (\texttt{gate\_proj}, \texttt{up\_proj}, and \texttt{down\_proj}). The maximum sequence length is set to 8,192 tokens for Mistral-7B-Instruct-v0.2 and 16,384 tokens for the other three models. All models are optimized using AdamW with a learning rate of $2 \times 10^{-4}$ and a cosine learning-rate scheduler. Each model is fine-tuned for three epochs.

\subsection{Vulnerability Inference}
\label{sec:inference_strategy}

During inference, we calculate a vulnerability probability instead of directly using the generated textual response. For each input sample, we extract the output logits corresponding to the ``Yes'' and ``No'' tokens. We then normalize these two logits to obtain the vulnerability probability:
\begin{equation}
P_{\mathrm{vul}} =
\frac{\exp(\operatorname{logit}(\mathrm{Yes}))}
{\exp(\operatorname{logit}(\mathrm{Yes}))
+\exp(\operatorname{logit}(\mathrm{No}))}.
\end{equation}

For each model, we select the classification threshold on the validation split of the ID benchmark and keep it fixed for evaluation. We vary $\tau$ from 0.01 to 0.90 with a step size of 0.01 and calculate the F1 score at each value. The threshold that produces the highest F1 score is selected as the optimal threshold $\tau^{*}$. The final prediction for a sample $x$ is defined as
\begin{equation}
\hat{y} =
\begin{cases}
1\;(\text{Vulnerable}), & \text{if } P_{\mathrm{vul}}(x) \geq \tau^{*}, \\
0\;(\text{Non-Vulnerable}), & \text{otherwise}.
\end{cases}
\end{equation}

After $\tau^{*}$ is selected, it remains fixed for the corresponding model. We use this same threshold to evaluate the model on the ID test set and all six OOD benchmarks. No threshold is searched for or adjusted on any test dataset. We apply the same procedure to all baseline models. Each baseline selects its own threshold on the same ID validation set and then uses that fixed threshold on the ID test set and all OOD benchmarks. This setting prevents test data from affecting threshold selection and ensures a consistent comparison across models and datasets. For the four fine-tuned GRAML models, the resulting thresholds are $\tau^{*}=0.41$ for CodeLlama, $0.44$ for Mistral, $0.53$ for DeepSeek, and $0.54$ for Qwen. We analyze how these validation thresholds relate to test performance in Section \ref{sec:threshold_robustness}.

\section{Evaluation}
\label{sec:evaluation}

We evaluate GRAML from several perspectives. We first describe the datasets and evaluation metrics, and then compare GRAML with three categories of baselines. Next, we evaluate its robustness under semantics-preserving code transformations. We further conduct ablation studies on the multi-task learning components and the graph-structured context extraction pipeline. Finally, we assess the effectiveness of ToT-VR and analyze threshold robustness and parameter sensitivity.

\subsection{Datasets}
\label{sec:datasets}

\begin{table}[htbp]
\centering
\caption{Statistics of the ID benchmark and OOD evaluation datasets.}
\label{tab:dataset_stats}
\resizebox{\columnwidth}{!}{%
\begin{tabular}{lcccc}
\toprule
\textbf{Dataset / Split} & \textbf{Total} & \textbf{Vulnerable} & \textbf{Non-Vuln} & \textbf{Vuln Ratio} \\
\midrule
\multicolumn{5}{l}{\textbf{\textit{In-Distribution Benchmark (ID)}}} \\
Training (multi-task) & 21,600 & 10,800 & 10,800 & 50.0\% \\
\hspace{3mm}\textit{- Detection} & \textit{19,137} & \textit{9,569} & \textit{9,568} & \textit{50.0\%} \\
\hspace{3mm}\textit{- Description} & \textit{823} & \textit{410} & \textit{413} & \textit{50.0\%} \\
\hspace{3mm}\textit{- Localization} & \textit{820} & \textit{410} & \textit{410} & \textit{50.0\%} \\
\hspace{3mm}\textit{- Assessment} & \textit{820} & \textit{410} & \textit{410} & \textit{50.0\%} \\
Validation & 2,310 & 1,155 & 1,155 & 50.0\% \\
ID Test & 2,311 & 1,155 & 1,156 & 50.0\% \\
\midrule
\multicolumn{5}{l}{\textbf{\textit{Out-of-Distribution Benchmarks (OOD)}}} \\
CVEfixes & 2,108 & 1,054 & 1,054 & 50.0\% \\
Devign & 936 & 468 & 468 & 50.0\% \\
DiverseVul & 766 & 383 & 383 & 50.0\% \\
PrimeVul & 548 & 274 & 274 & 50.0\% \\
Juliet (C/C++) & 1,576 & 788 & 788 & 50.0\% \\
ReVeal & 1,014 & 507 & 507 & 50.0\% \\
\bottomrule
\end{tabular}%
}
\end{table}

\begin{figure}[htbp]
    \centering
    \includegraphics[width=\linewidth]{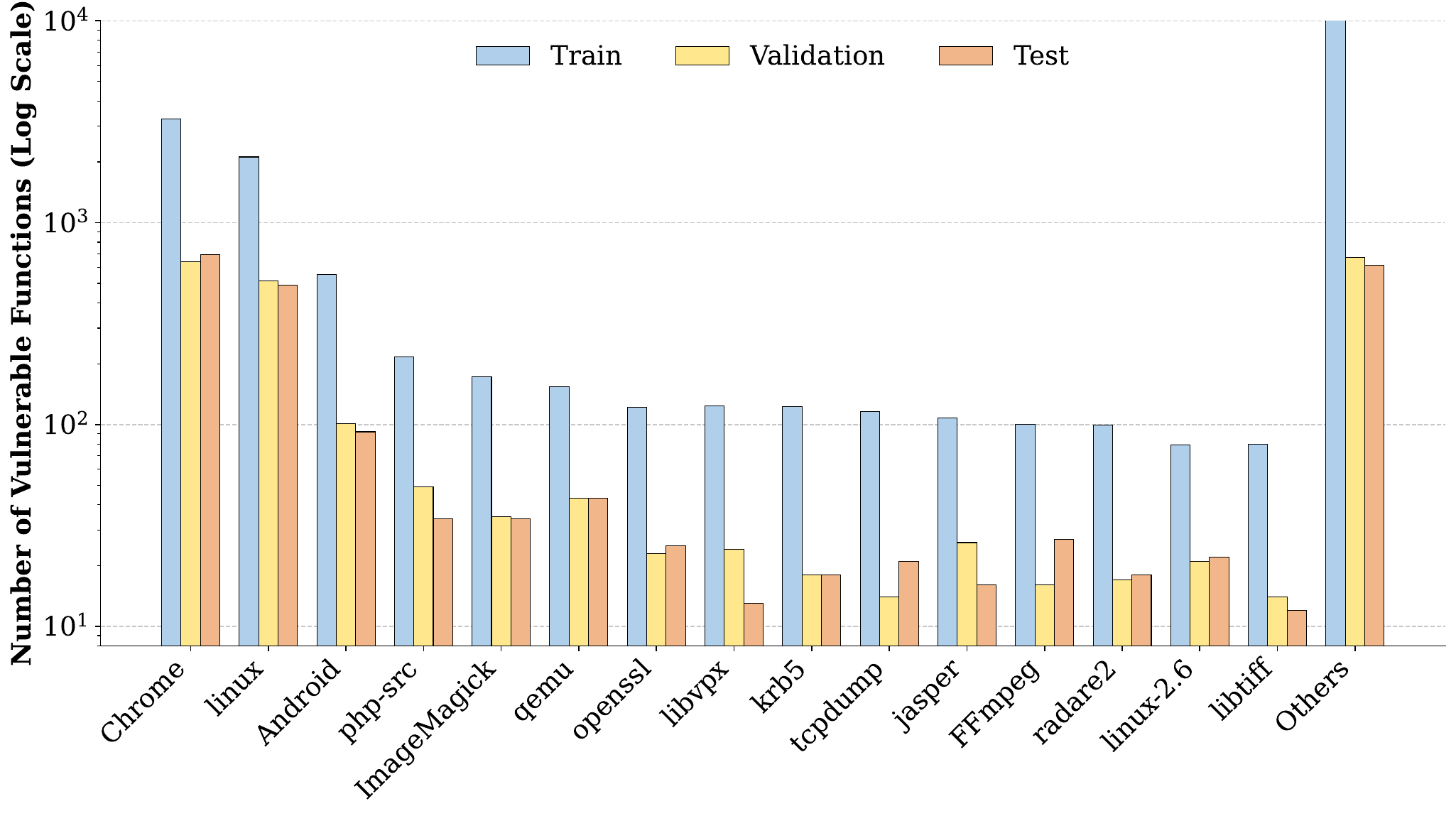}
    \caption{Project-level distribution of vulnerable functions in the training, validation, and test splits of the ID benchmark. The figure shows the 15 largest projects and the aggregated ``Others'' category.}
    \label{fig:project_distribution}
\end{figure}

We use one in-distribution benchmark and six out-of-distribution benchmarks. The ID benchmark is used for multi-task training, model selection, and internal testing. The OOD benchmarks are used only to evaluate cross-dataset generalization. Table~\ref{tab:dataset_stats} reports the statistics of all datasets. Figure~\ref{fig:project_distribution} presents the project-level distribution of the training, validation, and test splits in the ID benchmark.

\textbf{In-Distribution Benchmark (ID).}
We construct the ID benchmark from BigVul~\cite{fan2020ac}, DiverseVul~\cite{chen2023diversevul}, and additional vulnerable functions collected from CVE reports published between 2019 and 2026. After preprocessing and deduplication, the benchmark is divided into training, validation, and ID test sets. The three sets contain 21,600, 2,310, and 2,311 instances, respectively. Each set maintains an approximately balanced ratio between vulnerable and non-vulnerable instances.

The ID benchmark covers nearly 840 open-source projects. As shown in Figure~\ref{fig:project_distribution}, vulnerable functions in the training, validation, and ID test sets follow similar long-tailed project distributions. The figure reports the 15 largest projects and groups the remaining projects into the ``Others'' category.

The training set contains four types of task instances. Among the 21,600 training instances, 19,137 are used for vulnerability detection, 823 for vulnerability description, 820 for vulnerability localization, and 820 for vulnerability assessment. The validation and ID test sets contain only vulnerability-detection instances. Thus, all four tasks are used during fine-tuning, while model selection and internal evaluation focus on vulnerability detection.

\textbf{Out-of-Distribution Benchmarks (OOD).}
We use CVEfixes~\cite{bhandari2021cvefixes}, Devign~\cite{zhou2019devign}, DiverseVul~\cite{chen2023diversevul}, PrimeVul~\cite{ding2024vulnerability}, Juliet (C/C++)~\cite{boland2012juliet}, and ReVeal~\cite{chakraborty2021deep} as OOD benchmarks. These benchmarks are used only to evaluate vulnerability detection. Before evaluation, we remove samples that overlap with the ID benchmark through cross-dataset deduplication and filtering. As reported in Table~\ref{tab:dataset_stats}, each OOD benchmark contains approximately equal numbers of vulnerable and non-vulnerable instances. We use these benchmarks to evaluate whether the model can accurately distinguish vulnerable functions from non-vulnerable functions on external datasets that are not used during training.

\subsection{Evaluation Metrics}
\label{sec:evaluation_metrics}

We evaluate vulnerability detection using Precision, Recall, F1 score, Precision--Recall (PR) curves, and Average Precision (AP). We use the F1 score as the main evaluation metric because it considers both false alarms and missed vulnerabilities.

The F1 score is calculated from Precision and Recall as follows:
\begin{equation}
    F1 = 2 \times
    \frac{\mathrm{Precision} \times \mathrm{Recall}}
    {\mathrm{Precision} + \mathrm{Recall}}.
\end{equation}
Precision and Recall are defined as
\begin{equation}
    \mathrm{Precision} = \frac{TP}{TP + FP},
\end{equation}
and
\begin{equation}
    \mathrm{Recall} = \frac{TP}{TP + FN}.
\end{equation}
Here, $TP$ denotes the number of vulnerable samples correctly predicted as vulnerable. $FP$ denotes the number of non-vulnerable samples incorrectly predicted as vulnerable. $FN$ denotes the number of vulnerable samples incorrectly predicted as non-vulnerable. Precision therefore reflects the proportion of correct predictions among all samples predicted as vulnerable, while Recall reflects the proportion of detected samples among all truly vulnerable samples. A high F1 score requires both high Precision and high Recall.

We also report PR curves and AP as complementary metrics to the F1 score. A PR curve shows how Precision and Recall change under different decision thresholds. AP summarizes the PR curve into a single value and reflects the overall ranking quality of the predicted vulnerability scores. A higher AP indicates that vulnerable samples generally receive higher confidence scores than non-vulnerable samples. While the F1 score measures detection performance at one selected threshold, AP evaluates the model across all decision thresholds.

\begin{table*}[htbp]
\centering
\caption{Comparison of F1 scores (\%) on vulnerability detection. The best result in each column across all 15 models is highlighted in \textbf{bold}.}
\label{tab:results}

\fontsize{7.5pt}{8.5pt}\selectfont
\setlength{\tabcolsep}{0pt}
\renewcommand{\arraystretch}{1.03}

\begin{tabular*}{\textwidth}{@{\extracolsep{\fill}}l*{8}{c}@{}}
\toprule
\multirow{2}{*}{\textbf{Model}}
& \multicolumn{7}{c}{\textbf{Dataset}}
& \multirow{2}{*}{\textbf{Avg.}} \\
\cmidrule(lr){2-8}
& \textbf{CVEfixes}
& \textbf{Devign}
& \textbf{DiverseVul}
& \textbf{Juliet}
& \textbf{PrimeVul}
& \textbf{ReVeal}
& \textbf{ID Test}
& \\
\midrule

\multicolumn{9}{l}{\textbf{\textit{Encoder Baselines}}} \\
\midrule
CodeBERT
& 53.12 & 30.37 & 42.74 & 40.32 & 53.08 & 29.13 & 69.67 & 45.49 \\
GraphCodeBERT
& 53.95 & 39.95 & 46.98 & 55.33 & 58.06 & 52.78 & 69.95 & 53.86 \\
RoBERTa
& 58.92 & 39.95 & 46.41 & 56.73 & 47.41 & 64.94 & 67.03 & 54.48 \\
CodeBERTa
& 59.25 & 47.76 & 53.26 & 59.56 & 61.23 & 63.47 & 70.27 & 59.26 \\
GPT-2
& 50.04 & 26.75 & 37.48 & 52.03 & 39.32 & 47.00 & 67.33 & 45.71 \\
UniXcoder
& 57.59 & 42.55 & 49.21 & 51.65 & 50.33 & 49.71 & 70.86 & 53.13 \\
VulBERTa
& 51.45 & 16.80 & 46.36 & 8.90 & 56.58 & 15.03 & 69.35 & 37.78 \\
\midrule

\multicolumn{9}{l}{\textbf{\textit{Few-Shot LLMs}}} \\
\midrule
GLM-4.7
& 57.04 & 51.98 & 61.67 & 75.97 & 69.88 & 52.76 & 57.37 & 60.95 \\
DeepSeek-V4-Flash
& 65.00 & 60.24 & \textbf{67.46} & 68.52 & 70.42 & 62.45 & 61.31 & 65.06 \\
Qwen3-Coder-Next
& 30.00 & 35.26 & 32.87 & 76.70 & 40.93 & 22.98 & 58.37 & 42.44 \\
GPT-5
& 48.13 & 44.56 & 44.30 & 76.61 & 69.54 & 37.36 & 55.59 & 53.73 \\
\midrule

\multicolumn{9}{l}{\textbf{\textit{GRAML (Fine-tuned)}}} \\
\midrule
CodeLlama-7B
& 66.30 & 62.60 & 66.48 & \textbf{77.78}
& 70.15 & 67.03 & 70.58 & \textbf{68.70} \\
DeepSeek-6.7B
& \textbf{66.92} & \textbf{62.65} & 66.55 & 66.67
& \textbf{70.72} & 66.71 & 70.75 & 67.28 \\
Mistral-7B
& 65.19 & 61.41 & 64.81 & 65.02
& 69.72 & \textbf{68.36} & \textbf{72.19} & 66.67 \\
Qwen-7B
& 65.94 & 62.40 & 65.82 & 77.18
& 70.52 & 66.38 & 71.49 & 68.53 \\
\bottomrule
\end{tabular*}
\end{table*}

\begin{table}[htbp]
\centering
\caption{F1-score comparison (\%) between GRAML (GPT-4) and VulLLM.}
\label{tab:main_results_integrated}
\footnotesize
\renewcommand{\arraystretch}{1.16}

\begin{tabular*}{\columnwidth}{@{\extracolsep{\fill}}lcccc|cc@{}}
\toprule
\multirow{2}{*}{\textbf{Dataset}}
& \multicolumn{4}{c|}{\textbf{GRAML (GPT-4)}}
& \multicolumn{2}{c}{\textbf{VulLLM}} \\
\cmidrule(lr){2-5}\cmidrule(lr){6-7}
& \textbf{CL} & \textbf{DS} & \textbf{MS} & \textbf{QW}
& \textbf{L2} & \textbf{CL} \\
\midrule
CVEfixes   & 65.06 & \textbf{68.86} & 61.31 & 62.34 & 57.73 & 60.72 \\
Devign     & 62.19 & 62.49 & \textbf{63.97} & 61.93 & 60.43 & 62.88 \\
DiverseVul & 65.56 & \textbf{66.40} & 64.04 & 65.36 & 59.27 & 65.30 \\
Juliet     & \textbf{75.13} & 69.74 & 58.37 & 73.10 & 67.32 & 65.43 \\
PrimeVul   & 69.87 & 69.08 & \textbf{71.17} & 69.06 & 68.81 & 65.11 \\
ReVeal     & 66.79 & 66.35 & 70.37 & \textbf{71.68} & 57.26 & 65.88 \\
\midrule
\textbf{Average} & \textbf{67.43} & 67.15 & 64.87 & 67.25 & 61.80 & 64.22 \\
\bottomrule
\end{tabular*}
\vspace{1mm}
\begin{minipage}{\columnwidth}
\scriptsize
\textit{Note:} CL, DS, MS, QW, and L2 denote CodeLlama, DeepSeek,
Mistral, Qwen, and Llama2, respectively. Both frameworks use GPT-4
for vulnerability explanation generation. The best result in each
row is shown in bold. The average is computed over all six datasets.
\end{minipage}
\end{table}

\subsection{Performance Comparison}
\label{sec:performance_comparison}

This subsection evaluates the overall vulnerability detection performance of GRAML across multiple benchmarks. Table~\ref{tab:results} reports the F1 scores of 15 models on six OOD benchmarks and the ID test set. The average F1 score is calculated over all seven test sets.

\textbf{Comparison with Encoder Baselines.}
Among the seven encoder baselines, CodeBERTa achieves the highest average F1 score of 59.26\%. All four GRAML models outperform this result, with average F1 scores ranging from 66.67\% to 68.70\%. This corresponds to improvements of 7.41\% to 9.44\% over CodeBERTa ($66.67\%-59.26\%$ and $68.70\%-59.26\%$). Among the GRAML models, CodeLlama-7B achieves the highest overall average F1 score of 68.70\%. GRAML also outperforms all encoder baselines on every test set. The smallest improvement is 1.33\% on the ID test set, where Mistral-7B achieves 72.19\% compared with 70.86\% for UniXcoder. The largest improvement is 18.22\% on Juliet, where CodeLlama-7B achieves 77.78\% compared with 59.56\% for CodeBERTa. These results show that GRAML consistently outperforms the encoder baselines.

\textbf{Comparison with Few-Shot LLMs.}
Among the four Few-shot LLMs, DeepSeek-V4-Flash achieves the highest average F1 score of 65.06\%. The four fine-tuned GRAML models achieve average F1 scores ranging from 66.67\% to 68.70\%, improving over DeepSeek-V4-Flash by 1.61\% to 3.64\% ($66.67\%-65.06\%$ and $68.70\%-65.06\%$). Among the GRAML models, CodeLlama-7B achieves the highest average F1 score of 68.70\%. GRAML also achieves the best result on six of the seven test sets. The improvement ranges from 0.30\% to 10.88\%. On PrimeVul, DeepSeek-6.7B achieves 70.72\%, compared with 70.42\% for DeepSeek-V4-Flash. On the ID test set, Mistral-7B achieves 72.19\%, compared with 61.31\% for DeepSeek-V4-Flash. In addition, GPT-5 achieves an average F1 score of 53.73\% when directly used as a Few-shot detector, while GRAML with CodeLlama-7B achieves 68.70\%, an improvement of 14.97\% ($68.70\%-53.73\%$). These results show that the fine-tuned GRAML models outperform the evaluated Few-shot LLMs in average detection performance.

\textbf{Comparison with VulLLM.}
We further compare GRAML with VulLLM~\cite{du2024generalization}. To ensure a fair comparison, we follow the experimental protocol of VulLLM and re-run GRAML using GPT-4 instead of GPT-5, while keeping the remaining GRAML pipeline and training settings unchanged.

As shown in Table~\ref{tab:main_results_integrated}, all four GRAML models achieve higher average F1 scores than both VulLLM models. Their average F1 scores range from 64.87\% to 67.43\%, compared with 64.22\% for the strongest VulLLM variant. The improvements therefore range from 0.65\% to 3.21\%. GRAML with DeepSeek achieves the best results on CVEfixes and DiverseVul, Mistral performs best on Devign and PrimeVul, CodeLlama obtains the highest score on Juliet, and Qwen achieves the best result on ReVeal. Consequently, GRAML obtains the highest F1 score on all six datasets. These results show that GRAML is effective across different backbones and datasets.

\subsection{Robustness Evaluation}
\label{sec:Robustness}

To examine whether GRAML captures vulnerability semantics rather than relying on surface-level code patterns, we conduct robustness experiments following the adversarial evaluation framework of Risse et al.~\cite{risse2024uncovering}. We evaluate CodeLlama and Mistral as two representative GRAML backbones. CodeLlama is a code-specialized model and achieves the highest average F1 score in the main comparison, while Mistral is a general-purpose model. These two models therefore represent different backbone types and cover distinct performance patterns of GRAML.

Using these two representative backbones, we evaluate GRAML under a series of semantics-preserving code transformations. Table~\ref{tab:transformations} summarizes the seven transformations used in the robustness evaluation. These transformations are grouped into three categories. Semantic obfuscation (tf1--tf3) changes identifiers and comments without modifying program behavior. Long-context noise (tf4--tf5) inserts irrelevant content into the input. Structural restructuring (tf6--tf7) changes the organization of the code while preserving its semantics.

\begin{table*}[htbp]
\centering
\caption{Robustness evaluation in terms of F1 score (\%). The best result in each column is highlighted in \textbf{bold}.}
\label{tab:robustness}
\resizebox{\textwidth}{!}{%
\renewcommand{\arraystretch}{1.15}
\setlength{\tabcolsep}{2pt}
\begin{tabular}{@{}l*{21}{c}@{}}
\toprule
& \multicolumn{3}{c}{\textbf{CVEfixes}}
& \multicolumn{3}{c}{\textbf{Devign}}
& \multicolumn{3}{c}{\textbf{DiverseVul}}
& \multicolumn{3}{c}{\textbf{Juliet (C/C++)}}
& \multicolumn{3}{c}{\textbf{PrimeVul}}
& \multicolumn{3}{c}{\textbf{ReVeal}}
& \multicolumn{3}{c}{\textbf{ID Test}} \\
\cmidrule(lr){2-4}
\cmidrule(lr){5-7}
\cmidrule(lr){8-10}
\cmidrule(lr){11-13}
\cmidrule(lr){14-16}
\cmidrule(lr){17-19}
\cmidrule(lr){20-22}
\textbf{Model}
& Obfuscate & Noise & Structure
& Obfuscate & Noise & Structure
& Obfuscate & Noise & Structure
& Obfuscate & Noise & Structure
& Obfuscate & Noise & Structure
& Obfuscate & Noise & Structure
& Obfuscate & Noise & Structure \\
\midrule
\multicolumn{22}{l}{\textit{Encoder Baselines}} \\
\midrule
CodeBERT
&50.57&52.98&52.76
&28.00&29.88&30.10
&39.98&36.26&40.81
&19.64&0.00&0.00
&53.05&48.50&48.13
&20.35&29.11&24.63
&69.58&68.30&66.50 \\

CodeBERTa
&55.46&56.73&55.13
&50.41&42.05&44.16
&51.62&50.09&51.73
&25.12&47.78&51.25
&60.49&60.41&60.19
&62.61&62.85&62.38
&69.73&66.51&66.75 \\

GPT-2
&49.96&49.09&47.66
&22.17&24.53&24.41
&36.85&32.20&35.41
&\textbf{44.69}&49.77&41.01
&39.72&35.63&32.37
&46.02&45.56&43.18
&66.53&65.04&63.88 \\

GraphCodeBERT
&50.86&50.32&50.59
&36.91&37.58&33.66
&46.11&43.72&45.16
&1.01&0.00&0.00
&52.35&58.05&58.27
&51.09&50.32&49.19
&69.54&68.41&66.48 \\

RoBERTa
&55.48&55.06&54.54
&39.20&30.24&39.70
&42.92&44.36&43.23
&29.01&28.06&29.15
&43.17&47.35&46.76
&62.30&62.80&62.27
&67.06&64.90&65.25 \\

UniXcoder
&53.24&52.21&57.79
&37.20&40.68&41.30
&50.45&48.33&49.32
&8.76&19.28&0.25
&48.81&48.03&49.41
&49.53&49.41&48.06
&69.70&67.71&\textbf{68.89} \\
\midrule
\multicolumn{22}{l}{\textit{GRAML}} \\
\midrule
CodeLlama-7B
&\textbf{65.34}&\textbf{65.90}&\textbf{64.55}
&\textbf{61.63}&\textbf{62.56}&\textbf{62.56}
&\textbf{65.69}&\textbf{65.63}&\textbf{64.96}
&31.39&\textbf{67.17}&\textbf{69.00}
&\textbf{69.45}&\textbf{69.41}&\textbf{68.90}
&64.77&66.18&\textbf{66.09}
&68.61&68.93&68.75 \\

Mistral-7B
&59.76&59.22&58.62
&58.92&58.39&57.30
&61.80&61.20&60.96
&29.33&64.33&65.62
&69.27&67.04&67.07
&\textbf{65.65}&\textbf{66.80}&59.55
&\textbf{70.05}&\textbf{69.56}&65.14 \\
\bottomrule
\end{tabular}%
}
\end{table*}

\begin{figure*}[htbp]
    \centering
    \includegraphics[width=0.6\textwidth]{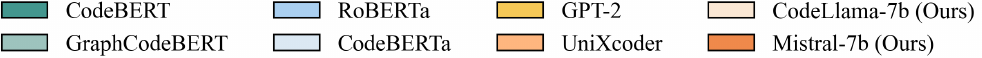}
    \\
    \subfloat[Semantic obfuscation]{
        \includegraphics[width=0.31\textwidth]{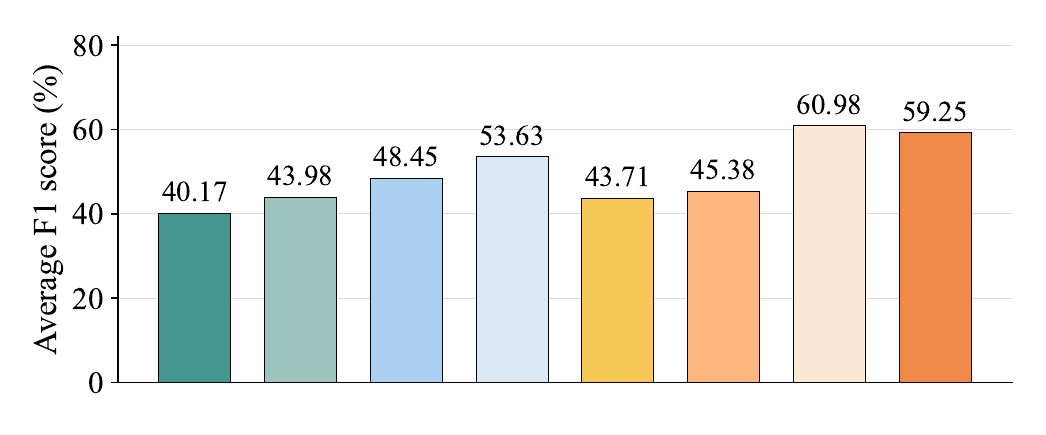}
        \label{fig:robustness_obfuscate}
    }
    \subfloat[Long-context noise]{
        \includegraphics[width=0.31\textwidth]{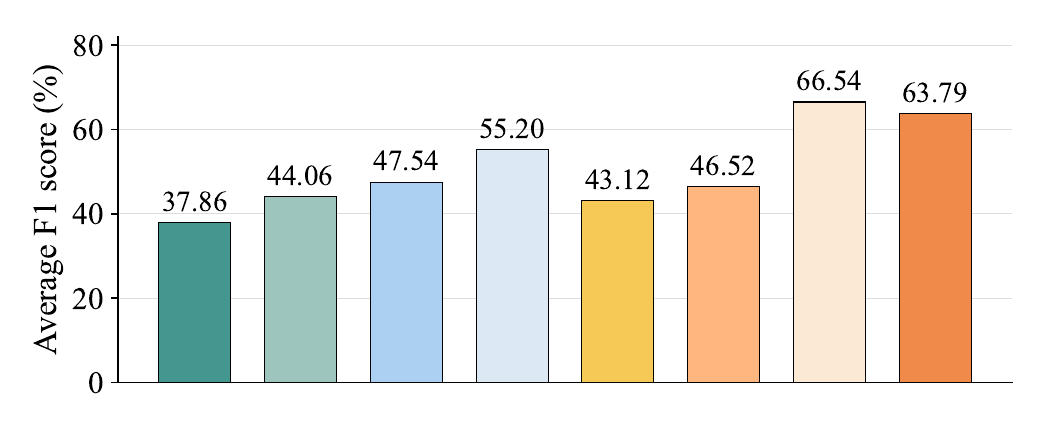}
        \label{fig:robustness_noise}
    }
    \subfloat[Structural restructuring]{
        \includegraphics[width=0.31\textwidth]{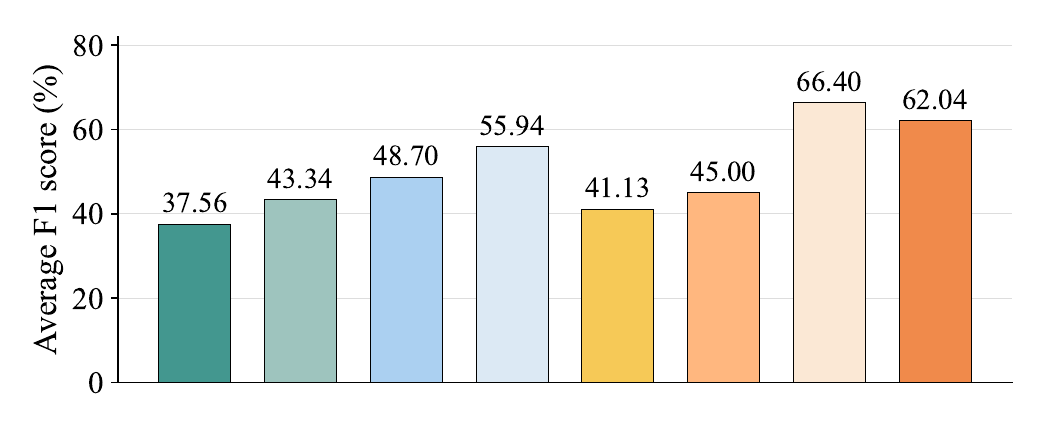}
        \label{fig:robustness_structure}
    }
    \caption{Average F1 scores across the seven datasets under semantic obfuscation, long-context noise, and structural restructuring.}
    \label{fig:robustness_chart}
\end{figure*}

\begin{table}[htbp]
\centering
\caption{Semantics-preserving transformations used for robustness evaluation.}
\label{tab:transformations}
\renewcommand{\arraystretch}{1.15}
\setlength{\tabcolsep}{5pt}

\begin{tabularx}{\columnwidth}{
    @{}
    >{\raggedright\arraybackslash}p{0.32\columnwidth}
    >{\centering\arraybackslash}p{0.08\columnwidth}
    X
    @{}
}
\toprule
\textbf{Category} & \textbf{ID} & \textbf{Transformation} \\
\midrule

\multirow{3}{*}{Semantic Obfuscation}
& tf1 & Parameter Renaming \\
& tf2 & Function-Name Renaming \\
& tf3 & Comment Removal \\

\midrule
\multirow{2}{*}{Long-context Noise}
& tf4 & Dead-Code Insertion \\
& tf5 & Irrelevant-Comment Insertion \\

\midrule
\multirow{2}{*}{\shortstack[l]{Structural\\Restructuring}}
& tf6 & Function Extraction \\
& tf7 & Void-Function Call Insertion \\

\bottomrule
\end{tabularx}
\end{table}

\textbf{Overall Results.}
Table~\ref{tab:robustness} reports the detailed F1 scores under the three perturbation categories, while Figure~\ref{fig:robustness_chart} presents the average results across the seven datasets. Under semantic obfuscation, long-context noise, and structural restructuring, CodeLlama achieves average F1 scores of 60.98\%, 66.54\%, and 66.40\%, respectively, while Mistral achieves 59.25\%, 63.79\%, and 62.04\%. Both GRAML backbones outperform the strongest encoder baseline in all three perturbation categories.

\textbf{Semantic Obfuscation.}
Under semantic obfuscation, the strongest encoder baseline, CodeBERTa, achieves an average F1 score of 53.63\%. In comparison, CodeLlama and Mistral achieve 60.98\% and 59.25\%, corresponding to improvements of 7.35\% ($60.98\%-53.63\%$) and 5.62\% ($59.25\%-53.63\%$), respectively. At the dataset level, CodeLlama obtains 65.34\% on CVEfixes and 61.63\% on Devign, compared with the strongest encoder results of 55.48\% and 50.41\%. GRAML also achieves the highest results on DiverseVul and ReVeal. These results indicate that both GRAML backbones remain effective when identifiers are renamed or comments are removed.

\textbf{Long-Context Noise.}
Under long-context noise, the strongest encoder baseline, CodeBERTa, achieves an average F1 score of 55.20\%. CodeLlama and Mistral achieve 66.54\% and 63.79\%, improving over CodeBERTa by 11.34\% ($66.54\%-55.20\%$) and 8.59\% ($63.79\%-55.20\%$), respectively. On Juliet, Mistral reaches 64.33\%, compared with 49.77\% for the strongest encoder baseline. On PrimeVul, CodeLlama obtains 69.41\%, while the strongest encoder result is 60.41\%. CodeLlama also achieves 65.90\% on CVEfixes, compared with 56.73\% for the strongest encoder baseline. These results show that GRAML can retain vulnerability-related evidence when dead code or irrelevant comments are inserted into the input.

\textbf{Structural Restructuring.}
Under structural restructuring, CodeBERTa achieves the highest average F1 score among the encoder baselines at 55.94\%. CodeLlama and Mistral achieve 66.40\% and 62.04\%, exceeding CodeBERTa by 10.46\% ($66.40\%-55.94\%$) and 6.10\% ($62.04\%-55.94\%$), respectively. On DiverseVul, CodeLlama achieves 64.96\%, compared with 51.73\% for the strongest encoder baseline. On Juliet, Mistral reaches 65.62\%, while the strongest encoder result is 51.25\%. CodeLlama also obtains 68.90\% on PrimeVul, exceeding the strongest encoder result of 60.19\%. These results show that GRAML remains stable when function boundaries and code organization are changed without altering program behavior.

\subsection{Ablation Study}
\label{sec:ablation}

To evaluate the contributions of the main components in GRAML, we conduct ablation studies from two perspectives. First, we examine the effectiveness of multi-task supervision by removing different auxiliary tasks from the unified four-task framework. Second, we evaluate the graph-evidence pipeline by changing or removing the CPG-guided supervision used for vulnerability description generation. Both studies are evaluated on the six external datasets.

\subsubsection{Effectiveness of Multi-Task Supervision}
\label{sec:ablation_multitask}

GRAML jointly optimizes the primary vulnerability Detection task together with three auxiliary tasks, namely Description, Localization, and Assessment. To determine whether these auxiliary tasks improve vulnerability detection and to measure the contribution of each task, we compare the full GRAML configuration with four ablated models. The \textbf{Detection Only} variant retains only the primary Detection task and removes all three auxiliary tasks. The abbreviation \textbf{w/o} means ``without.'' Therefore, \textbf{w/o Description}, \textbf{w/o Localization}, and \textbf{w/o Assessment} remove the corresponding auxiliary task individually while retaining the other tasks. We conduct this ablation study using CodeLlama and Mistral as two representative backbones that exhibit different performance characteristics in the main comparison. Table~\ref{tab:ablation} reports the F1 scores of the full configuration and the four ablated models on the six external datasets.

\begin{table*}[htbp]
\centering
\caption{Ablation results for multi-task supervision. The evaluated settings include \textbf{Detection Only}, \textbf{w/o Description}, \textbf{w/o Localization}, \textbf{w/o Assessment}, and the full GRAML configuration. Values in parentheses indicate $F1_{\mathrm{variant}}-F1_{\mathrm{GRAML}}$. The best and second-best results within each backbone are highlighted in \textbf{bold} and \underline{underlined}, respectively.}
\label{tab:ablation}

\setlength{\tabcolsep}{5mm}
\renewcommand{\arraystretch}{0.9}

\resizebox{\textwidth}{!}{%
\begin{tabular}{l|l|cccccc|c}
\toprule
\textbf{Model}
& \textbf{Setting}
& \textbf{CVEfixes}
& \textbf{Devign}
& \textbf{DiverseVul}
& \textbf{Juliet (C/C++)}
& \textbf{PrimeVul}
& \textbf{ReVeal}
& \textbf{Average} \\
\midrule

\multirow{5}{*}{CodeLlama-7B}
& Detection Only
& 58.26 \textcolor{red}{\scriptsize{(-8.04)}}
& 52.79 \textcolor{red}{\scriptsize{(-9.81)}}
& 62.94 \textcolor{red}{\scriptsize{(-3.54)}}
& 68.64 \textcolor{red}{\scriptsize{(-9.14)}}
& 66.29 \textcolor{red}{\scriptsize{(-3.86)}}
& 66.43 \textcolor{red}{\scriptsize{(-0.60)}}
& 62.56 \textcolor{red}{\scriptsize{(-5.83)}} \\

& w/o Description
& 59.80 \textcolor{red}{\scriptsize{(-6.50)}}
& 57.54 \textcolor{red}{\scriptsize{(-5.06)}}
& 64.53 \textcolor{red}{\scriptsize{(-1.95)}}
& 66.85 \textcolor{red}{\scriptsize{(-10.93)}}
& 68.08 \textcolor{red}{\scriptsize{(-2.07)}}
& 66.37 \textcolor{red}{\scriptsize{(-0.66)}}
& 63.86 \textcolor{red}{\scriptsize{(-4.53)}} \\

& w/o Localization
& 65.39 \textcolor{red}{\scriptsize{(-0.91)}}
& \underline{62.39} \textcolor{red}{\scriptsize{(-0.21)}}
& 65.59 \textcolor{red}{\scriptsize{(-0.89)}}
& 66.67 \textcolor{red}{\scriptsize{(-11.11)}}
& 67.95 \textcolor{red}{\scriptsize{(-2.20)}}
& \underline{66.85} \textcolor{red}{\scriptsize{(-0.18)}}
& 65.81 \textcolor{red}{\scriptsize{(-2.58)}} \\

& w/o Assessment
& \underline{65.61} \textcolor{red}{\scriptsize{(-0.69)}}
& 61.87 \textcolor{red}{\scriptsize{(-0.73)}}
& \underline{65.64} \textcolor{red}{\scriptsize{(-0.84)}}
& \underline{76.85} \textcolor{red}{\scriptsize{(-0.93)}}
& \underline{69.60} \textcolor{red}{\scriptsize{(-0.55)}}
& 66.58 \textcolor{red}{\scriptsize{(-0.45)}}
& \underline{67.69} \textcolor{red}{\scriptsize{(-0.70)}} \\

& Full GRAML
& \textbf{66.30}
& \textbf{62.60}
& \textbf{66.48}
& \textbf{77.78}
& \textbf{70.15}
& \textbf{67.03}
& \textbf{68.39} \\

\midrule

\multirow{5}{*}{Mistral-7B}
& Detection Only
& 54.91 \textcolor{red}{\scriptsize{(-10.28)}}
& 52.40 \textcolor{red}{\scriptsize{(-9.01)}}
& 58.57 \textcolor{red}{\scriptsize{(-6.24)}}
& 15.72 \textcolor{red}{\scriptsize{(-49.30)}}
& 66.67 \textcolor{red}{\scriptsize{(-3.05)}}
& 66.72 \textcolor{red}{\scriptsize{(-1.64)}}
& 52.50 \textcolor{red}{\scriptsize{(-13.25)}} \\

& w/o Description
& 58.24 \textcolor{red}{\scriptsize{(-6.95)}}
& 52.55 \textcolor{red}{\scriptsize{(-8.86)}}
& 61.16 \textcolor{red}{\scriptsize{(-3.65)}}
& 22.42 \textcolor{red}{\scriptsize{(-42.60)}}
& 69.07 \textcolor{red}{\scriptsize{(-0.65)}}
& \underline{66.86} \textcolor{red}{\scriptsize{(-1.50)}}
& 55.05 \textcolor{red}{\scriptsize{(-10.70)}} \\

& w/o Localization
& 60.85 \textcolor{red}{\scriptsize{(-4.34)}}
& 58.84 \textcolor{red}{\scriptsize{(-2.57)}}
& 62.19 \textcolor{red}{\scriptsize{(-2.62)}}
& 16.00 \textcolor{red}{\scriptsize{(-49.02)}}
& \textbf{70.78} \textcolor{natureGreenDark}{\scriptsize{(+1.06)}}
& 65.35 \textcolor{red}{\scriptsize{(-3.01)}}
& 55.67 \textcolor{red}{\scriptsize{(-10.08)}} \\

& w/o Assessment
& \underline{62.67} \textcolor{red}{\scriptsize{(-2.52)}}
& \underline{60.62} \textcolor{red}{\scriptsize{(-0.79)}}
& \underline{64.16} \textcolor{red}{\scriptsize{(-0.65)}}
& \underline{61.45} \textcolor{red}{\scriptsize{(-3.57)}}
& 69.33 \textcolor{red}{\scriptsize{(-0.39)}}
& 66.06 \textcolor{red}{\scriptsize{(-2.30)}}
& \underline{64.05} \textcolor{red}{\scriptsize{(-1.70)}} \\

& Full GRAML
& \textbf{65.19}
& \textbf{61.41}
& \textbf{64.81}
& \textbf{65.02}
& \underline{69.72}
& \textbf{68.36}
& \textbf{65.75} \\

\bottomrule
\end{tabular}%
}
\end{table*}

\begin{table*}[htbp]
\centering
\caption{Ablation results for the graph-evidence pipeline. Values in parentheses indicate $F1_{\mathrm{variant}}-F1_{\mathrm{GRAML}}$. The best result in each column within each backbone is highlighted in \textbf{bold}.}
\label{tab:ablation_graph}

\setlength{\tabcolsep}{5mm}
\renewcommand{\arraystretch}{0.9}

\resizebox{\textwidth}{!}{%
\begin{tabular}{l|l|cccccc|c}
\toprule
\textbf{Model}
& \textbf{Setting}
& \textbf{CVEfixes}
& \textbf{Devign}
& \textbf{DiverseVul}
& \textbf{Juliet (C/C++)}
& \textbf{PrimeVul}
& \textbf{ReVeal}
& \textbf{Average} \\
\midrule

\multirow{3}{*}{CodeLlama-7B}
& Full GRAML
& \textbf{66.30}
& \textbf{62.60}
& \textbf{66.48}
& \textbf{77.78}
& 70.15
& \textbf{67.03}
& \textbf{68.39} \\

& w/ Direct CPG
& 66.22 \textcolor{red}{\scriptsize{(-0.08)}}
& 61.80 \textcolor{red}{\scriptsize{(-0.80)}}
& 66.18 \textcolor{red}{\scriptsize{(-0.30)}}
& 66.67 \textcolor{red}{\scriptsize{(-11.11)}}
& \textbf{70.86} \textcolor{natureGreenDark}{\scriptsize{(+0.71)}}
& 66.91 \textcolor{red}{\scriptsize{(-0.12)}}
& 66.44 \textcolor{red}{\scriptsize{(-1.95)}} \\

& w/o Graph Evidence
& 64.49 \textcolor{red}{\scriptsize{(-1.81)}}
& 61.67 \textcolor{red}{\scriptsize{(-0.93)}}
& 64.67 \textcolor{red}{\scriptsize{(-1.81)}}
& 65.47 \textcolor{red}{\scriptsize{(-12.31)}}
& 68.52 \textcolor{red}{\scriptsize{(-1.63)}}
& 66.53 \textcolor{red}{\scriptsize{(-0.50)}}
& 65.23 \textcolor{red}{\scriptsize{(-3.16)}} \\

\midrule

\multirow{3}{*}{DeepSeek-6.7B}
& Full GRAML
& \textbf{66.92}
& \textbf{62.65}
& \textbf{66.55}
& \textbf{66.67}
& \textbf{70.72}
& \textbf{66.71}
& \textbf{66.70} \\

& w/ Direct CPG
& 66.84 \textcolor{red}{\scriptsize{(-0.08)}}
& 62.56 \textcolor{red}{\scriptsize{(-0.09)}}
& 66.49 \textcolor{red}{\scriptsize{(-0.06)}}
& 63.08 \textcolor{red}{\scriptsize{(-3.59)}}
& 69.45 \textcolor{red}{\scriptsize{(-1.27)}}
& 66.62 \textcolor{red}{\scriptsize{(-0.09)}}
& 65.84 \textcolor{red}{\scriptsize{(-0.86)}} \\

& w/o Graph Evidence
& 66.73 \textcolor{red}{\scriptsize{(-0.19)}}
& 62.60 \textcolor{red}{\scriptsize{(-0.05)}}
& 66.25 \textcolor{red}{\scriptsize{(-0.30)}}
& 62.45 \textcolor{red}{\scriptsize{(-4.22)}}
& 67.00 \textcolor{red}{\scriptsize{(-3.72)}}
& 66.67 \textcolor{red}{\scriptsize{(-0.04)}}
& 65.28 \textcolor{red}{\scriptsize{(-1.42)}} \\

\midrule

\multirow{3}{*}{Mistral-7B}
& Full GRAML
& \textbf{65.19}
& \textbf{61.41}
& \textbf{64.81}
& \textbf{65.02}
& \textbf{69.72}
& \textbf{68.36}
& \textbf{65.75} \\

& w/ Direct CPG
& 64.68 \textcolor{red}{\scriptsize{(-0.51)}}
& 61.24 \textcolor{red}{\scriptsize{(-0.17)}}
& 63.75 \textcolor{red}{\scriptsize{(-1.06)}}
& 63.67 \textcolor{red}{\scriptsize{(-1.35)}}
& 64.85 \textcolor{red}{\scriptsize{(-4.87)}}
& 66.36 \textcolor{red}{\scriptsize{(-2.00)}}
& 64.09 \textcolor{red}{\scriptsize{(-1.66)}} \\

& w/o Graph Evidence
& 50.57 \textcolor{red}{\scriptsize{(-14.62)}}
& 45.20 \textcolor{red}{\scriptsize{(-16.21)}}
& 50.44 \textcolor{red}{\scriptsize{(-14.37)}}
& 46.99 \textcolor{red}{\scriptsize{(-18.03)}}
& 54.81 \textcolor{red}{\scriptsize{(-14.91)}}
& 64.19 \textcolor{red}{\scriptsize{(-4.17)}}
& 52.03 \textcolor{red}{\scriptsize{(-13.72)}} \\

\midrule

\multirow{3}{*}{Qwen-7B}
& Full GRAML
& \textbf{65.94}
& 62.40
& \textbf{65.82}
& \textbf{77.18}
& \textbf{70.52}
& 66.38
& \textbf{68.04} \\

& w/ Direct CPG
& 64.82 \textcolor{red}{\scriptsize{(-1.12)}}
& 62.23 \textcolor{red}{\scriptsize{(-0.17)}}
& 64.89 \textcolor{red}{\scriptsize{(-0.93)}}
& 75.53 \textcolor{red}{\scriptsize{(-1.65)}}
& 70.14 \textcolor{red}{\scriptsize{(-0.38)}}
& 63.43 \textcolor{red}{\scriptsize{(-2.95)}}
& 66.84 \textcolor{red}{\scriptsize{(-1.20)}} \\

& w/o Graph Evidence
& 63.09 \textcolor{red}{\scriptsize{(-2.85)}}
& \textbf{62.55} \textcolor{natureGreenDark}{\scriptsize{(+0.15)}}
& 64.70 \textcolor{red}{\scriptsize{(-1.12)}}
& 66.55 \textcolor{red}{\scriptsize{(-10.63)}}
& 69.03 \textcolor{red}{\scriptsize{(-1.49)}}
& \textbf{66.67} \textcolor{natureGreenDark}{\scriptsize{(+0.29)}}
& 65.43 \textcolor{red}{\scriptsize{(-2.61)}} \\

\bottomrule
\end{tabular}%
}
\end{table*}

\textbf{Overall Effect of Multi-Task Supervision.}
The results show that training only on the Detection task is less effective than using the complete multi-task framework. Compared with Full GRAML, the Detection Only setting decreases the average F1 score by 5.83\% for CodeLlama ($68.39\%-62.56\%$) and 13.25\% for Mistral ($65.75\%-52.50\%$). The decline is observed across both backbones, showing that the three auxiliary tasks jointly provide useful supervision for vulnerability detection.

\textbf{Effect of the Description Task.}
Among the three individual auxiliary tasks, Description makes the largest contribution to average performance. Removing this task decreases the average F1 score by 4.53\% for CodeLlama ($68.39\%-63.86\%$) and 10.70\% for Mistral ($65.75\%-55.05\%$). These reductions are larger than those caused by removing Localization or Assessment individually. The effect is particularly evident on Juliet for Mistral, where the F1 score decreases from 65.02\% to 22.42\%. These results indicate that the Description task provides useful semantic supervision. It helps the models learn more information about the vulnerability than binary labels alone.

\textbf{Effects of the Localization and Assessment Tasks.}
Removing Localization decreases the average F1 score by 2.58\% for CodeLlama ($68.39\%-65.81\%$) and 10.08\% for Mistral ($65.75\%-55.67\%$). These results indicate that line-level supervision provides useful guidance for identifying vulnerability-related code regions and improves the detection task. Removing Assessment decreases the average F1 score by 0.70\% for CodeLlama ($68.39\%-67.69\%$) and 1.70\% for Mistral ($65.75\%-64.05\%$). Both decreases show that severity-related supervision provides useful information. Overall, these results show that Localization and Assessment provide useful supervision, and jointly optimizing all auxiliary tasks improves vulnerability detection.

\subsubsection{Effectiveness of Graph Evidence}
\label{sec:graph_evidence_ablation}

To evaluate the contribution of the graph-evidence pipeline, we extend this ablation study to all four GRAML backbones and construct two models that progressively modify the CPG-guided supervision.

\begin{itemize}
    \item \textbf{w/ Direct CPG} replaces the GPT-5-generated natural-language vulnerability descriptions with raw CPG serializations extracted by Joern. The Detection, Localization, and Assessment tasks remain unchanged. This variant examines whether directly using graph structures provides the same supervision as transforming them into natural-language vulnerability descriptions.

    \item \textbf{w/o Graph Evidence} removes the CPG-based graph evidence extraction while retaining the original GPT-5 and ToT-VR reasoning process. Specifically, the \texttt{[CPG Evidence]} block is removed, and GPT-5 directly analyzes the raw source code through the same ToT-VR procedure to generate vulnerability descriptions. All other tasks and training settings remain unchanged. This variant isolates the contribution of the extracted graph evidence while preserving the reasoning and description-generation process.
\end{itemize}

\textbf{Effect of Direct CPG Supervision.}
As shown in Table~\ref{tab:ablation_graph}, natural-language descriptions provide stronger supervision than raw CPG serializations. For CodeLlama, the average F1 score reaches 68.39\% with natural-language descriptions, compared with 66.44\% with raw CPG serializations. DeepSeek shows a similar pattern, with 66.70\% and 65.84\%, respectively. The same advantage appears for Mistral and Qwen. These gains suggest that a natural-language description turns the scattered data-flow and control-flow relations into a coherent, self-contained explanation, which the model can absorb more easily than a raw graph.

\textbf{Effect of Removing Graph Evidence.}
As shown in Table~\ref{tab:ablation_graph}, graph evidence contributes substantially to detection performance. For Mistral, the average F1 score reaches 65.75\% with graph evidence, but only 52.03\% without it. CodeLlama shows a similar trend, with 68.39\% and 65.23\%, respectively. The same improvement appears for DeepSeek and Qwen. These results suggest that the selected critical lines and typed relations give the reasoning branches concrete evidence about where a vulnerability is triggered and how it propagates, so the model can trace the causal path of the flaw instead of guessing from the source code alone.

\textbf{Dataset-Level Analysis.}
The contribution of graph evidence is largest on Juliet. On this dataset, Mistral reaches an F1 score of 65.02\% with graph evidence, compared with 46.99\% without it. CodeLlama shows the same trend, with 77.78\% and 65.47\%, respectively. The larger gains on this dataset suggest that Juliet's buffer-overflow samples hinge on precise data-flow relations, which the graph evidence captures explicitly and thereby helps the model identify these cases more reliably.

Overall, these results confirm the importance of both graph evidence and its representation format. GRAML benefits from extracting structural program information and converting it into natural-language vulnerability descriptions, which provide more effective supervision than raw CPG serializations or source code alone.

\subsection{Effectiveness of ToT-VR}
\label{sec:tot_effectiveness}

To evaluate the effectiveness of ToT-VR, we compare it with standard CoT for constructing vulnerability descriptions. Both methods use GPT-5, while the remaining data construction and training settings are kept unchanged. We separately construct two training sets using the descriptions generated by CoT and ToT-VR, and then fine-tune CodeLlama, Mistral, DeepSeek, and Qwen on each training set. The resulting models are evaluated on six external datasets. This controlled comparison isolates the effect of the reasoning strategy used for description generation.

As shown in Table~\ref{tab:cot_vs_tot}, ToT-VR consistently outperforms CoT across all four backbones and all six datasets. Across the four backbones, the overall average F1 score increases from 64.08\% with CoT to 67.22\% with ToT-VR, corresponding to an improvement of 3.14\% ($67.22\%-64.08\%$). At the model level, the average improvements range from 2.17\% for Mistral to 3.66\% for CodeLlama. More importantly, ToT-VR achieves higher F1 scores in all 24 model--dataset comparisons, showing that its benefit is consistent rather than limited to a particular backbone or benchmark.

The clearest gains are observed on Juliet, where ToT-VR increases the average F1 score across the four backbones from 67.23\% to 71.66\%. Similar improvements are also observed on CVEfixes and the other external datasets. These results show that ToT-VR provides more effective vulnerability-description supervision than standard CoT and consistently improves downstream vulnerability detection across different backbones.

Unlike standard CoT, which follows a single linear reasoning path, ToT-VR explores several candidate interpretations of the same function and keeps only those that survive the reflection and pruning stages. This search process makes ToT-VR less sensitive to early mistakes in the reasoning chain, since a branch that misreads the code can be dropped and replaced by a more plausible alternative. The descriptions produced in this way therefore align more closely with the actual vulnerability, and this alignment is what the downstream fine-tuning ultimately benefits from.

\begin{table*}[htbp]
    \centering
    \footnotesize
    \caption{F1-score comparison (\%) between CoT and ToT-VR. Values in parentheses indicate $F1_{\mathrm{ToT\mbox{-}VR}}-F1_{\mathrm{CoT}}$. The better result for each model--dataset pair is highlighted in \textbf{bold}.}
    \label{tab:cot_vs_tot}

    \begin{tabular*}{\textwidth}{@{\extracolsep{\fill}}lcccccccc@{}}
    \toprule
    \multirow{2}{*}{\textbf{Dataset}}
    & \multicolumn{2}{c}{\textbf{CodeLlama-7B}}
    & \multicolumn{2}{c}{\textbf{Mistral-7B}}
    & \multicolumn{2}{c}{\textbf{DeepSeek-6.7B}}
    & \multicolumn{2}{c}{\textbf{Qwen-7B}} \\
    \cmidrule(lr){2-3}
    \cmidrule(lr){4-5}
    \cmidrule(lr){6-7}
    \cmidrule(lr){8-9}
    & \textbf{CoT} & \textbf{ToT-VR}
    & \textbf{CoT} & \textbf{ToT-VR}
    & \textbf{CoT} & \textbf{ToT-VR}
    & \textbf{CoT} & \textbf{ToT-VR} \\
    \midrule

    CVEfixes
    & 63.68 & \textbf{66.30} \textcolor{natureGreenDark}{\tiny{(+2.62)}}
    & 63.74 & \textbf{65.19} \textcolor{natureGreenDark}{\tiny{(+1.45)}}
    & 59.43 & \textbf{66.92} \textcolor{natureGreenDark}{\tiny{(+7.49)}}
    & 62.58 & \textbf{65.94} \textcolor{natureGreenDark}{\tiny{(+3.36)}} \\

    Devign
    & 59.43 & \textbf{62.60} \textcolor{natureGreenDark}{\tiny{(+3.17)}}
    & 57.68 & \textbf{61.41} \textcolor{natureGreenDark}{\tiny{(+3.73)}}
    & 59.01 & \textbf{62.65} \textcolor{natureGreenDark}{\tiny{(+3.64)}}
    & 59.60 & \textbf{62.40} \textcolor{natureGreenDark}{\tiny{(+2.80)}} \\

    DiverseVul
    & 64.43 & \textbf{66.48} \textcolor{natureGreenDark}{\tiny{(+2.05)}}
    & 64.30 & \textbf{64.81} \textcolor{natureGreenDark}{\tiny{(+0.51)}}
    & 63.75 & \textbf{66.55} \textcolor{natureGreenDark}{\tiny{(+2.80)}}
    & 63.62 & \textbf{65.82} \textcolor{natureGreenDark}{\tiny{(+2.20)}} \\

    Juliet
    & 68.80 & \textbf{77.78} \textcolor{natureGreenDark}{\tiny{(+8.98)}}
    & 64.04 & \textbf{65.02} \textcolor{natureGreenDark}{\tiny{(+0.98)}}
    & 65.83 & \textbf{66.67} \textcolor{natureGreenDark}{\tiny{(+0.84)}}
    & 70.26 & \textbf{77.18} \textcolor{natureGreenDark}{\tiny{(+6.92)}} \\

    PrimeVul
    & 66.42 & \textbf{70.15} \textcolor{natureGreenDark}{\tiny{(+3.73)}}
    & 68.15 & \textbf{69.72} \textcolor{natureGreenDark}{\tiny{(+1.57)}}
    & 65.37 & \textbf{70.72} \textcolor{natureGreenDark}{\tiny{(+5.35)}}
    & 67.82 & \textbf{70.52} \textcolor{natureGreenDark}{\tiny{(+2.70)}} \\

    ReVeal
    & 65.60 & \textbf{67.03} \textcolor{natureGreenDark}{\tiny{(+1.43)}}
    & 63.58 & \textbf{68.36} \textcolor{natureGreenDark}{\tiny{(+4.78)}}
    & 65.32 & \textbf{66.71} \textcolor{natureGreenDark}{\tiny{(+1.39)}}
    & 65.46 & \textbf{66.38} \textcolor{natureGreenDark}{\tiny{(+0.92)}} \\

    \midrule
    \textbf{Average}
    & 64.73 & \textbf{68.39} \textcolor{natureGreenDark}{\tiny{(+3.66)}}
    & 63.58 & \textbf{65.75} \textcolor{natureGreenDark}{\tiny{(+2.17)}}
    & 63.12 & \textbf{66.70} \textcolor{natureGreenDark}{\tiny{(+3.58)}}
    & 64.89 & \textbf{68.04} \textcolor{natureGreenDark}{\tiny{(+3.15)}} \\
    \bottomrule
    \end{tabular*}
\end{table*}

\subsection{Threshold Robustness Analysis}
\label{sec:threshold_robustness}

The main experiments use decision thresholds selected on the ID validation set. To further examine how the classification thresholds affect the predictions of the four fine-tuned GRAML models, we analyze them from two complementary perspectives. The precision--recall (PR) curves evaluate how effectively the models rank vulnerable samples above non-vulnerable samples across all possible thresholds, while the threshold--F1 curves show how detection performance changes as the decision threshold varies.

\begin{figure}[htbp]
    \centering
    \includegraphics[width=\columnwidth]{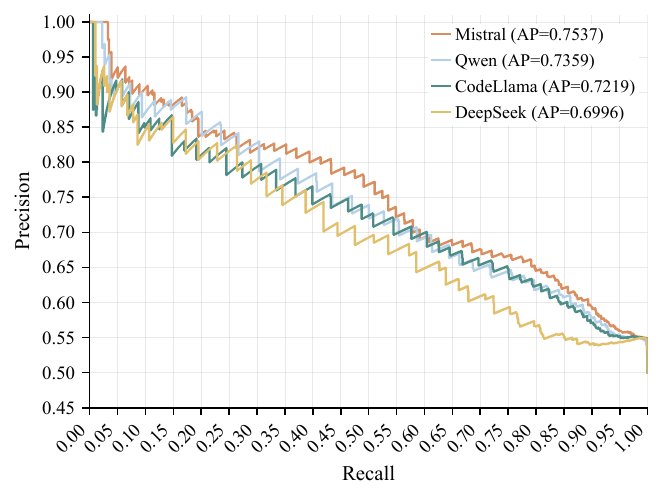}
    \caption{Precision--Recall curves of four fine-tuned GRAML models on the ID test set. AP denotes Average Precision and summarizes the area under the PR curve.}
    \label{fig:graml_pr_curves}
\end{figure}

As shown in Figure~\ref{fig:graml_pr_curves}, all four GRAML models achieve AP scores close to or above 0.70. Mistral obtains the highest AP of 0.7537, followed by Qwen, CodeLlama, and DeepSeek. The consistently strong AP scores indicate that the models produce meaningful vulnerability rankings across the full range of precision and recall, rather than relying only on predictions obtained at a single decision threshold.

\begin{figure}[htbp]
    \centering
    \includegraphics[width=\columnwidth]{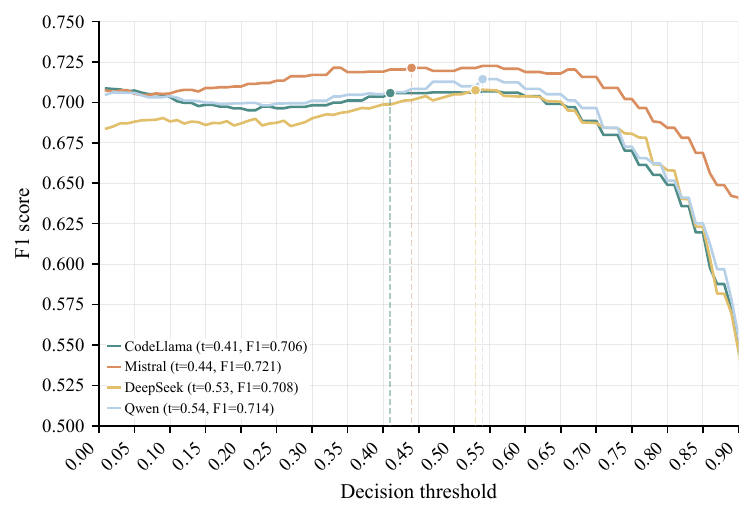}
    \caption{F1-score variations across decision thresholds for four fine-tuned GRAML models on the ID test set. The dashed lines indicate the thresholds selected on the ID validation set, and the markers show the corresponding F1 scores on the test sets.}
    \label{fig:f1_curves}
\end{figure}

To complement the PR-curve analysis, we further examine how the F1 score changes with the decision threshold. As shown in Figure~\ref{fig:f1_curves}, the thresholds selected on the ID validation set are $0.41$ for CodeLlama, $0.44$ for Mistral, $0.53$ for DeepSeek, and $0.54$ for Qwen, consistent with the thresholds reported in Section \ref{sec:inference_strategy}. Around these selected thresholds, the curves change gradually, and small threshold adjustments lead to limited variations in F1 score. Mistral achieves an F1 score of 72.19\% at its selected threshold, while the other three models obtain similar results. As the threshold increases further, the F1 scores of all four models decrease because fewer samples are classified as vulnerable, which reduces recall. Overall, the results show how the thresholds selected on the validation set correspond to the performance of the four models on the test sets.

\subsection{Parameter Sensitivity Analysis}
\label{sec:parameter_sensitivity}

We investigate the sensitivity of GRAML to the LoRA rank parameter $r$ while fixing the scaling factor $\alpha$ at 16. We vary $r$ over $\{4, 8, 16, 32, 64\}$ and evaluate CodeLlama-7B, Mistral-7B, DeepSeek-6.7B, and Qwen-7B. For each setting, we report the average F1 score over the six external datasets and the ID test set.

\begin{figure}[htbp]
    \centering
    \includegraphics[width=\columnwidth]{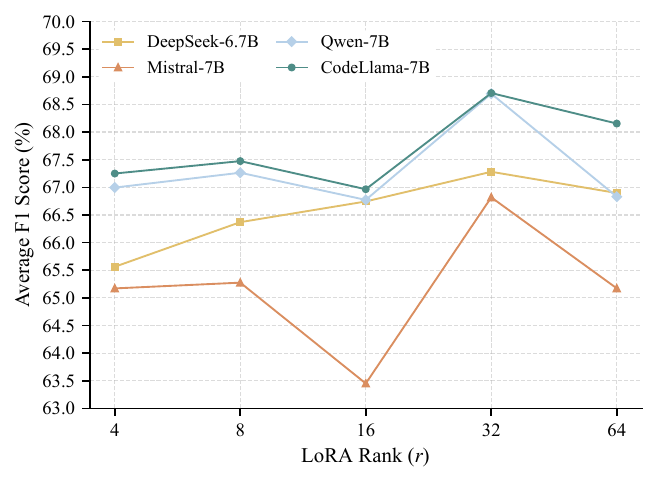}
    \caption{Average F1 scores across seven test datasets for four LLMs under different LoRA rank settings with $\alpha=16$.}
    \label{fig:parameter_sensitivity}
\end{figure}

As shown in Figure~\ref{fig:parameter_sensitivity}, all four backbones achieve their highest average F1 scores at $r=32$. CodeLlama and Qwen show similar performance trends and reach their best results at this setting. DeepSeek also performs best at $r=32$, while Mistral exhibits greater variation across different rank values but still achieves its highest score at $r=32$. These results indicate that $r=32$ provides the most effective and consistent configuration across the four backbones. We therefore use $r=32$ in the main experiments. Although the effect of the LoRA rank varies across models, the consistent optimum at $r=32$ shows that this setting provides a reliable choice for GRAML.

\section{Related Work}
\label{sec:related_work}

\subsection{Deep Learning for Vulnerability Detection}
Early vulnerability detection approaches modeled code as linear sequences using RNNs~\cite{li2018vuldeepecker} and LSTMs~\cite{li2021sysevr}. Ziems et al.~\cite{ziems2021security} further explored sequence-based models to improve vulnerability classification accuracy. Later studies adopted GNNs on CPGs to capture semantic dependencies more effectively~\cite{zhou2019devign} and extended these methods to incorporate additional structural information~\cite{chakraborty2021deep}. Subsequent works expanded these approaches to handle richer semantic patterns~\cite{qiu2024vulnerability} and more complex graph structures~\cite{zhang2023vulnerability}. Recent advancements incorporated causal reasoning~\cite{rahman2024causal}, dataflow analysis~\cite{steenhoek2024dataflow}, and graph denoising~\cite{cao2024snopy,cao2024coca}, alongside pre-trained models such as CodeBERT~\cite{feng2020codebert} and GraphCodeBERT~\cite{guo2020graphcodebert}, which learn general-purpose code representations~\cite{jiang2024stagedvulbert,li2025clever}.

Although LLMs such as CodeLlama~\cite{roziere2023code} and DeepSeek-Coder~\cite{guo2024deepseek} have demonstrated strong performance in vulnerability detection~\cite{ferrag2025securefalcon,lin2025large}, recent benchmarks suggest that such performance may be overestimated due to dataset noise and evaluation artifacts~\cite{ding2024vulnerability}. Furthermore, LLMs may still rely on superficial cues, such as variable names, rather than genuine semantic reasoning. As shown by~\cite{risse2024uncovering}, this reliance can cause performance drops under code obfuscation, highlighting the need for frameworks like GRAML that guide LLMs to capture the underlying causal logic of vulnerabilities.

\subsection{Structure-Aware Program Analysis}
To address the limitations of treating code purely as text, researchers have gradually incorporated structural information into LLM-based vulnerability analysis. Early approaches combined static structure with dynamic symbolic traces to enrich contextual information beyond linear tokens~\cite{wang2024combining}, while subsequent research leveraged CPGs to expose control and data flow dependencies difficult to capture from linear code sequences~\cite{lekssays2025llmxcpg,wang2024llmdfa}. For example, LLMxCPG~\cite{lekssays2025llmxcpg} uses CPG-derived context to assist LLM-based vulnerability analysis. In contrast, GRAML uses CPG evidence as a training-time constraint and distills it into natural-language supervision, so that the fine-tuned detector does not require graph evidence at inference. More recent methods have explored instruction-level graphs and LLM-enhanced code representations to improve generalization in vulnerability prediction~\cite{wen2025towards}. Benchmarks indicate that models still struggle to interpret long-range data flow contexts~\cite{li2025sv}, and graph-based methods can be sensitive to noisy structural dependencies and long-range relations~\cite{peng2025keep}. Static-analysis-assisted LLM frameworks also use program analysis results to augment reasoning about vulnerabilities~\cite{li2025iris}. In addition, hybrid fusion of language models and graph neural networks has been explored to jointly model structural and semantic features for vulnerability classification~\cite{liu2025vul}, and lightweight directed graph constructions capture multi-scale lexical co-occurrence and fine-grained structural dependencies~\cite{liu2026litevul}.

Recent work has further advanced structure-aware vulnerability detection by integrating fine-grained program graph analysis. For example, GraphFVD~\cite{shao2025graphfvd} extracts detailed structural features from program graphs and leverages hierarchical embeddings to capture syntax and semantic information relevant to vulnerabilities. Similarly, Zou et al.~\cite{zou2025code} propose a vulnerability detection framework based on an augmented program dependency graph and optimized CodeBERT, extending traditional PDG structures to better preserve semantic and structural code information. Motivated by these observations, our approach leverages structural dependencies to guide vulnerability analysis, allowing GPT-5 to explore execution-aware interpretations grounded in the code’s control and data flow.

\subsection{Reasoning and Multi-Task Learning}
Beyond structural integration, explicit reasoning and multi-task supervision are important for improving the reliability of LLM-based vulnerability analysis. CoT prompting~\cite{wei2022chain} improves reasoning by eliciting intermediate reasoning paths, while Self-Consistency~\cite{wang2022self} aggregates multiple reasoning results to enhance stability. Tree of Thoughts (ToT) further extends this process into a structured search over multiple candidate reasoning branches~\cite{yao2023tree,long2023large}. In the context of security research, explicit reasoning frameworks have been shown to provide detailed insights into vulnerability causes and improve interpretability in automated analysis~\cite{kim2025logs}. Recent studies have also adopted semantic analysis for vulnerability repair and root cause analysis~\cite{nong2025appatch}. Evaluations across multiple tasks suggest that jointly considering multiple objectives yields more robust performance than optimizing single tasks independently~\cite{yin2024multitask}.

Meanwhile, fine-tuning methods that are parameter efficient, such as LoRA~\cite{hu2022lora} and QLoRA~\cite{dettmers2023qlora}, enable efficient adaptation of large foundation models to downstream tasks without retraining the full network. Recent security studies have explored explanation generation~\cite{mao2025explainable}, which helps models produce interpretable reasoning paths for vulnerabilities, and task collaboration~\cite{du2024generalization}, where multiple related tasks are trained jointly to improve overall model consistency. Task collaboration has additionally been shown to benefit multi-objective supervision~\cite{wadhwa2024core}, enabling models to align performance across detection, localization, and assessment objectives. Motivated by these studies, GRAML uses structural evidence from graphs to guide ToT-based vulnerability reasoning and generate supervision across detection, localization, and assessment tasks, unifying structured reasoning with multi-task learning.

\section{Discussion}
\label{sec:discussion}

Code structure is important for helping LLMs go beyond surface-level patterns. GRAML focuses on vulnerability detection, which determines whether a given function is vulnerable. Before constructing the multi-task training dataset, GRAML uses CPG-based structural evidence to guide GPT-5 in generating vulnerability descriptions. These descriptions serve as the target outputs of the Description task and help encode the causal logic behind vulnerabilities. GRAML then builds a unified four-task dataset around the main Detection task. The Description task explains why the vulnerability occurs, the Localization task identifies where the vulnerable lines are, and the Assessment task estimates how severe the vulnerability is. Together, these tasks help the model learn not only whether a function is vulnerable, but also why it is vulnerable, where the problem occurs, and how serious the impact is. By learning from these structurally guided descriptions and auxiliary labels, GRAML reduces the model's reliance on shallow cues such as variable names, improves detection performance on ID data, and generalizes better to OOD benchmarks.

This study focuses on C/C++ vulnerabilities as an initial step toward structure-aware vulnerability detection. Future work will extend GRAML to multiple programming languages, including Python, Java, and other widely used languages, and will incorporate more diverse vulnerability datasets. We will also integrate additional static analysis tools to improve the extraction of structural program information and explore the use of GRAML for automated program repair.

\section{Conclusion}
\label{sec:conclusion}

In this paper, we present GRAML, a vulnerability detection framework that combines reliability-aware CPG evidence extraction, ToT-VR-based vulnerability description generation, and multi-task instruction tuning. GRAML selects critical source lines and typed control-flow and data-flow relations, which guide GPT-5 in generating natural-language descriptions of the vulnerability cause, the affected code regions, and the potential security consequences. These descriptions are jointly learned with Detection, Localization, and Assessment tasks, while the fine-tuned models require only source code during inference. Experimental results on the ID test set and six external datasets show that the best GRAML model achieves an average F1 score of 68.70\%, outperforming state-of-the-art baselines by up to 30.92\%. The ablation results further demonstrate that converting graph evidence into natural-language supervision is more effective than directly using raw CPG serializations or removing graph evidence. GRAML also maintains consistent performance under the evaluated semantics-preserving code transformations. Overall, the results demonstrate the effectiveness of integrating static program analysis, structured reasoning, and multi-task supervision for LLM-based vulnerability detection.

\appendices

\section{CPG Extraction and Context Construction}
\label{sec:appendix_cpg_context}

When generating vulnerability descriptions, using the entire function as input may introduce irrelevant information and obscure vulnerability-related code. In contrast, using only the vulnerable lines may remove the semantic context required to understand the vulnerability. To address these issues, we use Joern to construct CPGs and extract structural information from the source code. The resulting graph information is subsequently processed to support vulnerability description generation with GPT-5.

Applying Joern to vulnerability datasets is challenging because many samples are incomplete code snippets rather than complete compilable projects. To improve parsing stability and obtain usable CPGs, we address the following issues:

\begin{itemize}
    \item \textbf{Challenge 1: Parsing C++ Snippets.}

    \textit{Problem:} Many vulnerability datasets contain only isolated functions. Joern may fail to parse C++ scope resolution operators such as \texttt{::} when the surrounding class or namespace definitions are unavailable.

    \textit{Solution:} We normalize the source code before parsing by replacing \texttt{::} with \texttt{\_\_}. This preprocessing step reduces parsing failures and allows Joern to construct the corresponding abstract syntax tree and CPG.

    \item \textbf{Challenge 2: Dense and Redundant Graphs.}

    \textit{Problem:} The raw DOT graphs generated with the \texttt{--repr all} option contain numerous nodes and edges. Some nodes represent metadata, such as \texttt{METHOD} declarations, while some edges connect nodes located on the same source line. These elements increase graph size without providing useful line-level context.

    \textit{Solution:} We retain only nodes associated with valid source-line numbers and remove metadata nodes. We also discard edges whose source and target nodes belong to the same line. The remaining graph contains four relation types: Control Flow (\textsc{CFG}), Control Dependence (\textsc{CDG}), Data Dependence (\texttt{REACHING\_DEF}), and Function Call (\textsc{CALL}). These relations are then mapped to line-level structural representations.

    \item \textbf{Challenge 3: Residual Files during Batch Processing.}

    \textit{Problem:} Joern may raise internal errors when processing malformed or complex code snippets. Such failures can leave residual files, including \texttt{replay\_pid*.log}, which consume storage space and interfere with large-scale batch processing.

    \textit{Solution:} We execute Joern through a multiprocessing wrapper that isolates each parsing process, suppresses unnecessary input and output streams, and removes residual files after execution. This design allows the pipeline to continue processing subsequent samples when individual parsing failures occur.
\end{itemize}

\section{Practical Implementation of ToT-VR}
\label{sec:appendix_tot_vr_implementation}

As described in the main manuscript, ToT-VR generates vulnerability descriptions through five sequential stages. We implement this process using the GPT-5 API and generate multiple candidate descriptions for each vulnerable sample before selecting the final result.

ToT-VR introduces additional processing time because each sample requires five sequential API calls. Generating descriptions for 410 vulnerable samples in one iteration takes approximately three hours. Unlike few-shot generation, which requires only one API call for each sample, ToT-VR performs five consecutive reasoning stages. Because the prompt at Step $N$ depends on the output of Step $N-1$, the five calls for the same sample cannot be executed in parallel.

The generated descriptions may vary across runs because of the nondeterministic nature of LLM generation. We therefore use multiple iterations rather than relying on a single generation result. Specifically, we run the complete ToT-VR pipeline five times for each of the 410 samples, producing five candidate vulnerability descriptions for every sample.

We evaluate the candidate descriptions according to the following four criteria:

\begin{itemize}
    \item \textbf{Root Cause Accuracy.} The description should correctly identify the primary cause of the vulnerability and remain consistent with the corresponding CWE or CVE information.

    \item \textbf{Contextual Completeness.} The description should cover the relevant dependency path, affected code, and conditions required to trigger the vulnerability.

    \item \textbf{Impact Clarity.} The description should clearly explain the potential security impact, such as denial of service or remote code execution.

    \item \textbf{Factual Consistency.} The description should remain consistent with the given source code and should not introduce nonexistent variables, functions, or program behaviors.
\end{itemize}

We manually review the five candidate descriptions according to these criteria and select the description with the best overall assessment for each sample. Although this process introduces additional manual effort, it helps exclude descriptions containing unsupported or inaccurate information and improves the quality of the Description-task training samples.

\section{Fine-Tuning Implementation}
\label{sec:appendix_fine_tuning}

\subsection{Task Instruction Prompts}
\label{app:task_instruction_prompts}

The four tasks in the multi-task fine-tuning framework are represented using a unified instruction format. Each training instance is composed of a task-specific instruction, a source-code input, and a target response. The task instruction describes what the model is expected to do and the exact format of the expected output. To make the fine-tuning process fully reproducible, we list the complete instruction for each task in Table~\ref{tab:task_prompts}. All four tasks share the same global system prompt: ``You are a security assistant for vulnerability detection, assessment, location and description.'' During fine-tuning, the instruction and source-code tokens are masked from the loss computation, and the loss is calculated only on the target-response tokens.

\begin{table}[htbp]
\centering
\caption{Complete Task Instructions Used During Multi-Task Fine-Tuning}
\label{tab:task_prompts}
\footnotesize
\renewcommand{\arraystretch}{1.2}
\begin{tabularx}{\columnwidth}{>{\hsize=.14\hsize\raggedright\arraybackslash}X >{\hsize=.86\hsize\raggedright\arraybackslash}X}
\toprule
\textbf{Task} & \textbf{Instruction} \\
\midrule
\textbf{Detection} & You are a security code vulnerability analyzer. Your task is to carefully analyze the provided code snippet. Note that the provided code snippet might not be complete, but it has all the important context. Your output must be EXACTLY ONE WORD: If you detect any potential security vulnerability in the specified code segment, return: Yes. If the code segment appears to be secure and free from obvious vulnerabilities, return: No. Consider common vulnerability types such as buffer overflows, improper input validation, integer overflow, memory corruption potential, double free, and use after free. Your response must be either ``Yes'' or ``No'' --- no additional explanation. \\
\midrule
\textbf{Description} & You are a security code vulnerability analyzer. Your task is to analyze the following code and determine whether it contains vulnerabilities. Write the final vulnerability explanation as one coherent paragraph in professional security-analysis style that clearly states what the vulnerability is, why it occurs as the root cause, how it can be triggered, how the internal logic or data or control flow leads to the flaw, and what impact an attacker may achieve. The output must be a single paragraph of 3 to 5 sentences, must not contain any parentheses, and may quote small relevant portions of the source code without including line numbers or bracket references. If you determine that there is no vulnerability, only output ``There is no vulnerability.'' \\
\midrule
\textbf{Location} & You are a security code vulnerability analyzer. Your task is to carefully analyze the following C code and identify the exact locations of potential vulnerabilities. If any vulnerabilities exist, output the zero-based vulnerable line indices separated by commas. Consider the vulnerable statements, operations, and code regions that directly contribute to the security issue. Output format example: 1,4,7. If no vulnerabilities are found, only output ``N/A''. Do not provide any additional explanation. \\

\midrule
\textbf{Assessment} & You are a security code vulnerability analyzer. Your task is to analyze the following C code snippet and provide the CVSS v3.0 severity rating based on the identified vulnerability. Use the following mapping: High = 2, Medium = 1, Low = 0. Consider the potential impact and exploitability of the vulnerability when assigning the severity level. Output only one digit, either 2 or 1 or 0, and do not output any other text. If you determine that there is no vulnerability, only output ``N/A''. \\

\bottomrule
\end{tabularx}
\end{table}

\subsection{Prompt Templates for Target LLMs}
\label{app:target_llm_prompts}

During supervised fine-tuning (SFT), the instruction tuning data must be formatted according to the native chat template of each LLM. These templates define how system prompts, user instructions, input code, and assistant responses are separated by special tokens. If the template does not match the official format, the model may fail to parse different parts of the input correctly during training. Therefore, we use the official conversational template for each target model. Table~\ref{tab:prompt_templates} summarizes the formatting structures used in our Unsloth based training script.

\begin{table}[htbp]
\centering
\caption{Official Chat Templates Used During Supervised Fine Tuning}
\label{tab:prompt_templates}
\footnotesize
\renewcommand{\arraystretch}{1.2}
\begin{tabularx}{\columnwidth}{>{\hsize=.12\hsize\raggedright\arraybackslash}X >{\hsize=.88\hsize\raggedright\arraybackslash}X}
\toprule
\textbf{Model} & \textbf{Native Prompt Template Structure} \\
\midrule
Mistral
& \texttt{<s>[INST] <\!<SYS>\!><system><\!</SYS>\!> \#\# Input: <input> [/INST]} \\
\midrule
DeepSeek-Coder
& \texttt{<system> \#\#\# Instruction: <instr> <input> \#\#\# Response:} \\
\midrule
CodeLlama
& \texttt{<system> \#\#\# Instruction: <instr> \#\#\# Input: <input> \#\#\# Response:} \\
\midrule
Qwen
& \texttt{<|im\_start|>system <system><|im\_end|> <|im\_start|>user <instr> <input><|im\_end|> <|im\_start|>assistant} \\
\bottomrule
\end{tabularx}
\end{table}

We also implement a custom completion-only data collator. For each model, the collator identifies the token sequence that marks the beginning of the assistant response according to the corresponding template in Table~\ref{tab:prompt_templates}. It then sets the labels of all preceding tokens to \texttt{-100}. Consequently, the training loss is computed only on the assistant response, which contains the vulnerability detection result and explanation, rather than on the system prompt, user instruction, or input code.

\subsection{Sequence Length Analysis}

The maximum sequence length used in fine-tuning is determined by both the length of training samples and the available GPU memory. As described above, we set the maximum sequence length to 8,192 for Mistral and 16,384 for DeepSeek, CodeLlama, and Qwen. This subsection explains the rationale behind this setting.

Each training input contains the task instruction, source code, extracted graph aware context, and target response. Therefore, short sequence lengths may truncate not only the source code but also the structural context or the expected output. To examine this issue, we analyze the token length distribution of the training dataset. Since source code is often fragmented and may produce more tokens than natural language, we estimate token length conservatively. As shown in Table~\ref{tab:seq_length_distribution}, 91.04\% of the training samples are within 2,048 tokens. However, 2.97\% of the samples exceed 4,096 tokens, and 0.93\% exceed 8,192 tokens. Some samples are even longer than 16,384 tokens.

\begin{table}[htbp]
\centering
\caption{Estimated Token Length Distribution of the Training Dataset}
\label{tab:seq_length_distribution}
\small
\renewcommand{\arraystretch}{1.15}
\begin{tabularx}{\columnwidth}{>{\hsize=1.2\hsize}X >{\hsize=0.9\hsize\raggedleft\arraybackslash}X >{\hsize=0.9\hsize\raggedleft\arraybackslash}X}
\toprule
\textbf{Token Length} & \textbf{Count} & \textbf{\%} \\
\midrule
$\le 2,048$ & 16,808 & 91.04 \\
$2,049 - 4,096$ & 1,106 & 5.99 \\
$4,097 - 8,192$ & 376 & 2.04 \\
$8,193 - 16,384$ & 135 & 0.73 \\
$> 16,384$ & 37 & 0.20 \\
\bottomrule
\end{tabularx}
\end{table}

Based on this distribution, commonly used sequence lengths such as 2,048 or 4,096 would cut off many training samples. When a sample is cut off, the model may lose vulnerability related code lines, graph aware context, or response tokens. Therefore, we use longer context windows when possible. We set the maximum sequence length to 8,192 for Mistral and 16,384 for DeepSeek, CodeLlama, and Qwen. A small number of samples longer than 16,384 tokens are still cut off. This setting reduces information loss while keeping training feasible on our hardware.

\section{Prompt Templates for ToT-VR}
\label{sec:appendix_tot_prompts}

Table~\ref{tab:tot_prompts} presents the complete prompt templates used in the five stages of ToT-VR. All stages share the same system prompt, while each stage uses a separate user prompt and output format.

\begin{table*}[htbp]
\centering
\caption{Prompt templates used in the proposed Tree-of-Thought-guided Vulnerability Reasoning (ToT-VR) process.}
\label{tab:tot_prompts}
\scriptsize
\renewcommand{\arraystretch}{1.15}
\begin{tabularx}{\textwidth}{p{0.18\textwidth}X}
\toprule
\textbf{Prompt Type / Step} & \textbf{Prompt Template} \\
\midrule

\textbf{Shared System Prompt} &
You are a security analysis assistant specialized in source-code vulnerability reasoning. Follow the instructions for each step and use only the information provided in the current prompt. When multiple candidate branches are required, label them as Branch1, Branch2, and Branch3. When refining, pruning, or verifying branches, base the decision on the provided code and evidence. Follow the required output format and do not include unrelated content. \\
\midrule

\textbf{Step 1:\newline Branch Initialization} &
\textbf{Input:} Source code and up to 20 critical lines. Each critical line contains a physical line number and its source-code text. \newline
\textbf{User Prompt:} \newline
Analyze the following code and construct exactly three candidate reasoning branches for the potential vulnerability. Each branch should contain: (1) a possible vulnerability type, and (2) a brief explanation of its trigger point or causal mechanism. \newline
\textbf{Code:} \texttt{<FUNCTION\_CODE>} \newline
\textbf{Critical Lines:} \texttt{<CRITICAL\_LINE\_RECORDS>} \newline
\textbf{Output Format:} \newline
Branch1: [possible vulnerability type] -- [brief explanation] \newline
Branch2: [possible vulnerability type] -- [brief explanation] \newline
Branch3: [possible vulnerability type] -- [brief explanation] \\
\midrule

\textbf{Step 2:\newline Branch Reflection and Pruning} &
\textbf{Input:} Source code and the candidate branches generated in Step 1. \newline
\textbf{User Prompt:} \newline
Review the candidate branches based on the source code. Check whether each branch is consistent with the code semantics. Revise unclear branches when necessary and prune branches that are not supported by the code. Retain only the branches that remain plausible. \newline
\textbf{Code:} \texttt{<FUNCTION\_CODE>} \newline
\textbf{Candidate Branches:} \texttt{<BRANCHES\_FROM\_STEP\_1>} \newline
\textbf{Output Format:} \newline
Retained Branches: [list of retained branches] \newline
Pruned Branches: [list of pruned branches] \newline
Reason: [brief explanation of the pruning result] \\
\midrule

\textbf{Step 3:\newline Branch Refinement} &
\textbf{Input:} Retained branches from Step 2, selected vulnerable lines, and typed line relations. \newline
\textbf{User Prompt:} \newline
Refine the retained branches using the following structural evidence. Use the selected vulnerable lines and typed line relations to clarify the trigger point, dependency path, and causal explanation of each branch. Prune any branch that is not supported by this evidence. \newline
\textbf{Retained Branches:} \texttt{<RETAINED\_BRANCHES\_FROM\_STEP\_2>} \newline
\textbf{Selected Vulnerable Lines:} \texttt{<SELECTED\_VULNERABLE\_LINES>} \newline
\textbf{Typed Line Relations:} \texttt{<TYPED\_LINE\_RELATIONS>} \newline
\textbf{Output Format:} \newline
Refined Branches: [updated retained branches] \newline
Pruned Branches: [list of additionally pruned branches] \newline
Reason: [brief explanation of the refinement result] \\
\midrule

\textbf{Step 4:\newline Branch Verification} &
\textbf{Input:} Refined branches from Step 3 and the official CVE description. \newline
\textbf{User Prompt:} \newline
Compare the refined branches with the official CVE description. Determine which branch is most consistent with the known vulnerability type, cause, trigger condition, and impact. Select the best-supported branch as the verified branch. If no branch is sufficiently supported, report that the branches should be reconsidered. \newline
\textbf{Refined Branches:} \texttt{<REFINED\_BRANCHES\_FROM\_STEP\_3>} \newline
\textbf{Official CVE Description:} \texttt{<CVE\_DESCRIPTION>} \newline
\textbf{Output Format:} \newline
Verified Branch: [selected branch or none] \newline
Verification Result: [consistent / reconsider] \newline
Reason: [brief explanation of the verification result] \\
\midrule

\textbf{Step 5:\newline Description Synthesis} &
\textbf{Input:} Verified branch from Step 4. \newline
\textbf{User Prompt:} \newline
Generate a final vulnerability description based only on the verified branch. Write one coherent paragraph that explains the vulnerability cause, the affected code regions, and the potential security consequences. Do not include branch labels, bullet points, or intermediate reasoning. \newline
\textbf{Verified Branch:} \texttt{<VERIFIED\_BRANCH>} \newline
\textbf{Output Format:} A single paragraph containing only the final vulnerability description. \\
\bottomrule
\end{tabularx}
\end{table*}

\begin{figure*}[htbp]
    \centering
    \includegraphics[width=0.98\textwidth]{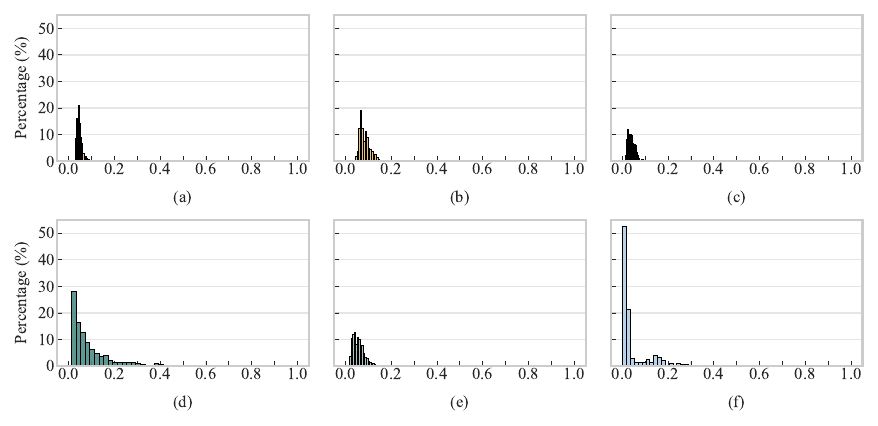}
    \caption{Confidence collapse across baseline models under different perturbations on the Juliet dataset. Subfigures correspond to (a) CodeBERT under Noise, (b) CodeBERT under Structure, (c) GraphCodeBERT under Noise, (d) GraphCodeBERT under Obfuscate, (e) GraphCodeBERT under Structure, and (f) UniXcoder under Structure.}
    \label{fig:confidence_collapse}
\end{figure*}

\section{Robustness Transformations}
\label{sec:appendix_robustness}

As introduced in the robustness evaluation section of the main manuscript, we use semantics-preserving code transformations to examine whether the fine-tuned GRAML models capture vulnerability semantics rather than relying on surface-level code patterns. Following the adversarial evaluation framework of Risse et al.~\cite{risse2024uncovering}, we apply seven transformations grouped into three categories: Semantic Obfuscation, Long-Context Noise, and Structural Restructuring. The implementation of each transformation is described below.

\subsection{Semantic Obfuscation}

This category examines whether the models rely on readable parameter names, function names, or comments. Such lexical information may be unavailable or intentionally modified in real-world vulnerable code. Semantic Obfuscation contains the following three transformations.

\begin{itemize}
    \item \textbf{Parameter Renaming (tf1):}
    We parse each function signature and extract its parameter names, such as \texttt{buf} from \texttt{void func(char* buf)}. We then replace each parameter name in the function body with a randomly generated two-letter string, such as \texttt{xk}. Identifier-boundary checks are applied to avoid replacing substrings within other identifiers.

    \item \textbf{Function-Name Renaming (tf2):}
    We identify the target function name and replace it with a randomly generated two-letter string. This transformation removes the semantic information conveyed by descriptive function names. For example, \texttt{test\_buffer\_overflow} may be replaced with \texttt{qz}, requiring the model to analyze the function body rather than its name.

    \item \textbf{Comment Removal (tf3):}
    We use a regular-expression-based parser to remove both single-line comments marked by \texttt{//} and multi-line comments enclosed by \texttt{/* ... */}. This transformation removes natural-language hints and requires the model to rely on executable code.
\end{itemize}

\subsection{Long-Context Noise}

This category inserts irrelevant content that increases the input length without changing the behavior of the original program. It examines whether the models can retain vulnerability-related information when the input contains unrelated code or text. Long-Context Noise contains the following two transformations.

\begin{itemize}
    \item \textbf{Dead-Code Insertion (tf4):}
    We append an unexecuted helper function, \texttt{void helpfunc()}, to each code sample. The function contains a \texttt{while(false)} block with 300 consecutive \texttt{break} statements. Because the condition is always false and the helper function is not called, the inserted code does not affect the behavior of the original function but substantially increases the input length.

    \item \textbf{Irrelevant-Comment Insertion (tf5):}
    We append a multi-line comment block enclosed by \texttt{/* ... */} to each code sample. The comment repeats the word \texttt{break} 50 times. The inserted text does not affect program behavior but introduces repetitive lexical noise into the input.
\end{itemize}

\subsection{Structural Restructuring}

This category modifies the organization of the source code while preserving its original behavior. It examines whether the models remain effective when vulnerability-related statements are reorganized or preceded by additional function calls. Structural Restructuring contains the following two transformations.

\begin{itemize}
    \item \textbf{Function Extraction (tf6):}
    We locate the boundaries of the original function body through brace matching using \texttt{\{ ... \}}. The body is moved into a new helper function named \texttt{helper\_func}, while the original parameters are preserved. The original function is then modified to invoke \texttt{helper\_func} with the corresponding arguments. This transformation changes the location and organization of the vulnerability-related logic without modifying its behavior.

    \item \textbf{Void-Function Call Insertion (tf7):}
    We define an empty function named \texttt{void help\_func()} containing 150 \texttt{return} statements and append it to the code sample. A call to \texttt{help\_func()} is inserted immediately after the opening brace of the target function. This transformation introduces an additional benign interprocedural call before the original function logic.
\end{itemize}

These seven transformations correspond directly to tf1--tf7 in Table~\ref{tab:transformations} of the main manuscript.

\section{Baseline Failure Analysis}
\label{sec:appendix_failure_analysis}

As reported in the main paper, several baseline encoder models, including CodeBERT, GraphCodeBERT, and UniXcoder, show clear performance drops on the Juliet dataset under different adversarial perturbations. In some cases, their F1 scores decrease to nearly zero. To examine whether these results are caused by evaluation artifacts or pipeline failures, we conduct a sanity check and failure analysis.

The analysis shows that the near zero F1 scores are mainly caused by tokenization truncation and confidence collapse. Sequence based models usually impose a maximum sequence length, such as 512 tokens. The perturbations increase the code length and cause all perturbed samples in the analyzed subsets to exceed this limit. The average length is more than 860 tokens. As a result, the actual vulnerable lines are truncated and fall outside the input range of the models.

After the vulnerable lines are truncated, the models lose important semantic information for prediction. As shown in Figure~\ref{fig:confidence_collapse}, the predicted probabilities of true vulnerable samples are mostly below the standard 0.5 decision threshold. The models, therefore, classify most samples as benign and produce zero true positives. These results indicate that the zero F1 scores are not caused by evaluation errors. Instead, they reflect the sensitivity of sequence based encoder models to sequence inflation.

\bibliographystyle{IEEEtran}
\bibliography{main}

\end{document}